\documentclass[useAMS,usenatbib]{mnras}

\usepackage[dvipdfmx]{color} 
\usepackage{graphicx}
\usepackage{color}
\usepackage{xcolor}
\usepackage{amssymb}
\title[GK Per 2015 dwarf nova covered by Suzaku]{
Suzaku X-ray observation of the GK Persei dwarf nova outburst in 2015
}
\author[
T. Yuasa et al.
]{
Takayuki Yuasa$^{1}$\thanks{E-mail: takayuki.yuasa@riken.jp},
Takayuki Hayashi$^{2,3}$, and
Manabu Ishida$^{4,5}$\\
$^{1}$Nishina Center for Accelerator-based Science, RIKEN,
2-1 Hirosawa, Wako, Saitama 351-0198, Japan\\
$^{2}$Goddard Space Flight Center, National Aeronautics and Space Administration, Greenbelt, MD 20771, USA\\
$^{3}$Department of Physics, Faculty of Science, Nagoya University, Furo-Cho, Chikusa-ku, Nagoya 464-8602, Japan\\
$^{4}$The Institute of Space and Astronautical Science,
Japan Aerospace Exploration Agency, \\
~3-1-1 Yoshinodai, Chuo-ku, Sagamihara 252-5210, Japan\\
$^{5}$Department of Physics, Tokyo Metropolitan University,
1-1 Minami-Osawa, Hachioji, Tokyo 192-0397, Japan
}

\begin{document}


\newcommand{\red}{\textcolor{red}}
\newcommand{\green}{\textcolor{green}}
\newcommand{\blue}{\textcolor{blue}}

\newcommand{\suzaku}{{\it Suzaku} }
\newcommand{\integral}{{\it INTEGRAL} }
\newcommand{\rxte}{{\it RXTE} }
\newcommand{\asca}{{\it ASCA} }
\newcommand{\rosat}{{\it ROSAT} }
\newcommand{\einstein}{{\it Einstein} }
\newcommand{\chandra}{{\it Chandra} }
\newcommand{\xmm}{{\it XMM-Newton} }
\newcommand{\newton}{{\it XMM-Newton} }
\newcommand{\swift}{{\it Swift} }
\newcommand{\ginga}{{\it Ginga} }
\newcommand{\sax}{{\it BeppoSAX} }
\newcommand{\ah}{{\it ASTRO-H} }
\newcommand{\astroh}{{\it ASTRO-H} }
\newcommand{\nustar}{{\it NuSTAR} }

\newcommand{\kms}{\mathrm{km}~\mathrm{s}^{-1}}
\newcommand{\erg}{\mathrm{erg}}
\newcommand{\eV}{\mathrm{eV}}
\newcommand{\persec}{\mathrm{s}^{-1}}
\newcommand{\arcminsq}{\mathrm{arcmin}^{-2}}
\newcommand{\cmsq}{\mathrm{cm}^{-2}}
\newcommand{\cmsqpersec}{\mathrm{cm}^{-2}~\persec}
\newcommand{\cmcubed}{\mathrm{cm}^{-3}}
\newcommand{\countss}{\mathrm{counts}\ \mathrm{s}^{-1}}
\newcommand{\ergs}{\mathrm{erg}\ \mathrm{s}^{-1}}
\newcommand{\ergcms}{\mathrm{erg}\ \mathrm{cm}^{-2}\ \mathrm{s}^{-1}}
\newcommand{\ergcmsdeg}{\mathrm{erg}\ \mathrm{cm}^{-2}\ \mathrm{s}^{-1}~\mathrm{deg}^{-2}}
\newcommand{\ergcmssr}{\mathrm{erg}\ \mathrm{cm}^{-2}\ \mathrm{s}^{-1}\ \mathrm{sr}^{-1}}
\newcommand{\gcms}{\mathrm{g}\ \mathrm{cm}^{-2}\ \mathrm{s}^{-1}}

\newcommand{\mwd}{M_{\mathrm{WD}}}
\newcommand{\rwd}{R_{\mathrm{WD}}}
\newcommand{\zfe}{Z_{\mathrm{Fe}}}
\newcommand{\msun}{M_\odot}
\newcommand{\zsun}{Z_\odot}
\newcommand{\nel}{n_{\mathrm{e}}}
\newcommand{\mel}{m_{\mathrm{e}}}
\newcommand{\kB}{k_{\mathrm{B}}}
\newcommand{\zs}{z_{\mathrm{s}}}
\newcommand{\Ts}{T_{\mathrm{s}}}
\newcommand{\nH}{n_{\mathrm{H}}}
\newcommand{\chisq}{\chi^2}
\newcommand{\chisqv}{\chi^{2}_\nu}

\newcommand{\feka}{Fe~K$\alpha$~}

\newcommand{\vsgr}{V1223~Sgr}


\ifdefined \aj \else
\newcommand\aj{{AJ}}%
\newcommand\actaa{{Acta Astron.}}%
\newcommand\araa{{ARA\&A}}%
\newcommand\apj{{ApJ}}%
\newcommand\apjl{{ApJ}}%
\newcommand\apjs{{ApJS}}%
\newcommand\ao{{Appl.~Opt.}}%
\newcommand\apss{{Ap\&SS}}%
\newcommand\aap{{A\&A}}%
\newcommand\aapr{{A\&A~Rev.}}%
\newcommand\aaps{{A\&AS}}%
\newcommand\azh{{AZh}}%
\newcommand\baas{{BAAS}}%
\newcommand\caa{{Chinese Astron. Astrophys.}}%
\newcommand\cjaa{{Chinese J. Astron. Astrophys.}}%
\newcommand\icarus{{Icarus}}%
\newcommand\jcap{{J. Cosmology Astropart. Phys.}}%
\newcommand\jrasc{{JRASC}}%
\newcommand\memras{{MmRAS}}%
\newcommand\mnras{{MNRAS}}%
\newcommand\na{{New A}}%
\newcommand\nar{{New A Rev.}}%
\newcommand\pra{{Phys.~Rev.~A}}%
\newcommand\prb{{Phys.~Rev.~B}}%
\newcommand\prc{{Phys.~Rev.~C}}%
\newcommand\prd{{Phys.~Rev.~D}}%
\newcommand\pre{{Phys.~Rev.~E}}%
\newcommand\prl{{Phys.~Rev.~Lett.}}%
\newcommand\pasa{{PASA}}%
\newcommand\pasp{{PASP}}%
\newcommand\pasj{{PASJ}}%
\newcommand\qjras{{QJRAS}}%
\newcommand\rmxaa{{Rev. Mexicana Astron. Astrofis.}}%
\newcommand\skytel{{S\&T}}%
\newcommand\solphys{{Sol.~Phys.}}%
\newcommand\sovast{{Soviet~Ast.}}%
\newcommand\ssr{{Space~Sci.~Rev.}}%
\newcommand\zap{{ZAp}}%
\newcommand\nat{{Nature}}%
\newcommand\iaucirc{{IAU~Circ.}}%
\newcommand\aplett{{Astrophys.~Lett.}}%
\newcommand\apspr{{Astrophys.~Space~Phys.~Res.}}%
\newcommand\bain{{Bull.~Astron.~Inst.~Netherlands}}%
\newcommand\fcp{{Fund.~Cosmic~Phys.}}%
\newcommand\gca{{Geochim.~Cosmochim.~Acta}}%
\newcommand\grl{{Geophys.~Res.~Lett.}}%
\newcommand\jcp{{J.~Chem.~Phys.}}%
\newcommand\jgr{{J.~Geophys.~Res.}}%
\newcommand\jqsrt{{J.~Quant.~Spec.~Radiat.~Transf.}}%
\newcommand\memsai{{Mem.~Soc.~Astron.~Italiana}}%
\newcommand\nphysa{{Nucl.~Phys.~A}}%
\newcommand\physrep{{Phys.~Rep.}}%
\newcommand\physscr{{Phys.~Scr}}%
\newcommand\planss{{Planet.~Space~Sci.}}%
\newcommand\procspie{{Proc.~SPIE}}%
\fi

\newcommand{\fittingEnergyBand}{$4.5-8$}

\date{Accepted 2016 March 24. Received 2016 March 19; in original form 2016 January 31}

\pagerange{\pageref{firstpage}--\pageref{lastpage}} \pubyear{2015}

\maketitle

\label{firstpage}

\begin{abstract}
The intermediate polar GK Per exhibited a dwarf nova outburst in March-April 2015.
\suzaku X-ray telescope serendipitously captured the onset of the outburst
during its pre-scheduled pointing observation spanning four days.
In this paper, we present temporal and spectral analysis results of
this outburst, together with those from archival data of quiescent obtained in 2009 and 2014.
Our temporal analysis confirmed previously reported spin modulation of X-ray count rates in outburst with a WD spin period of $P_{\mathrm{WD}}=351.4\pm0.5$~s. 
The modulation is also detected in the hard X-ray band ($16-60$~keV), and spectral modeling of the absorption suggests obscuration by a dense absorption with a line-of-sight column density of $N_\mathrm{H}>10^{23}~\cmsq$.
A complex time evolution of spin modulation profiles is seen; the spin minimum phase shifts
from phase $\sim0.25$ in the first half of the observation to $\sim0.65$ in the second one,
and the pulse shape significantly changes epoch by epoch.
Spectral fitting in the Fe K$\alpha$ band revealed an increase of 
the fluorescent line equivalent width, from $\sim80$~eV (quiescent) to $\sim140$~eV (outburst).
The equivalent widths of He-like and H-like Fe K$\alpha$ are consistent with being
constant at $\sim40$~eV in the two states. Broad-band spectral fitting in
the $2-60$~keV band resulted in a sub-solar Fe abundance of $\sim0.1~Z_\odot$ and
the maximum plasma temperature $kT_{\mathrm{max}}\sim50-60$~keV when the isobaric cooling flow model was applied.
Based on the very small temperature change against a $6-7$-times increased
accretion rate, the accretion geometry in early outburst is discussed.
\end{abstract}


\section{Introduction}
GK Persei, hereafter GK Per, is a cataclysmic variable (CV) system that
hosts a magnetized white dwarf (WD) and is observable as an intermediate polar.
This source underwent a classical nova explosion in 1901 \citep{williams1901,hale1901} and also exhibits dwarf nova outbursts about every 3~years each lasting
about 50 days (e.g. \citealt{simon2002}). Although GK Per is a magnetic
system that has magnetically channeled accretion within its magnetosphere,
it is believed that the dwarf nova outburst occurs due to increased
mass transfer caused by instability within an accretion disk circulating
around the magnetosphere of the WD (e.g. \citealt{nogamietal2002}). 
\citet{morales-ruedaetal2002gkper} measured an absorption spectrum of the companion
star, and reported mass ratio of the companion and the WD as $q=M_\mathrm{K}/M_\mathrm{WD}=0.55\pm0.21$, and constrained the WD mass as $M_\mathrm{WD}>0.87\pm0.24~M_\odot$. \citet{suleimanovetal2005} estimated a WD mass
based on X-ray spectral fitting as $M_\mathrm{WD}=0.59\pm0.05~M_\odot$ noting that the value may be underestimated by $\sim20\%$ due to reduced magnetospheric radius in outburst in which their \rxte data were taken. \citet{hachisukato2007gkper} derived the best-fit WD mass of $1.15\pm0.1~M_\odot$ based on the classical nova light curve modeling. 

Because of its unique characteristics (i.e. magnetic CV with classical nova/dwarf nova activies), GK Per has been deeply studied in broad 
wavelength; see \citealt{nogamietal2002} and \citealt{takeietal2015} for extensive
collections of previous publications.
Among many studies, key milestones in X-ray, that are relevant to the present paper,
include detections of 
(1) the spin period of 351~s using {\it EXOSAT} \citep{watsonetal1985}, 
(2) the complex absorption structure and Fe K$\alpha$ emission line structure using \ginga and \asca \citep{ishidaetal1992, ezukaishida1999},
(3) the potential red-ward wing in Fe K$\alpha$ line using \chandra HETG \citep{hellierandmukai2004fek}, and
(4) emission lines from lighter elements such as O and N using \xmm RGS \citep{vrielmannetal2005}.
Studies of the X-ray emitting expanding shell which was produced by the 1901 nova suggested presence of non-equilibrium ionization collisional plasma in the nova remnant \citep{balman2005,takeietal2015}.

Dwarf nova outbursts of GK Per provide us with interesting opportunities to study
disk instability \citep{kimetal1992,nogamietal2002} and a disk-magnetosphere
interaction in a changing mass accretion rate ($\sim\times10-20$ compared to
that of quiescence; \citealt{ishidaetal1992}).
In the present paper, we analyze X-ray observation data of the latest dwarf nova
occurred in March-April 2015 to investigate the accretion geometry and the structure
of the post-shock accretion region such as the maximum temperature and 
the magnetospheric radius in outburst.
We also compare results of these analyses with those obtained in quiescent when available.

\section[]{Observation and data reduction}\label{sec:obs}

\begin{figure*}
\includegraphics[width=0.85\hsize]{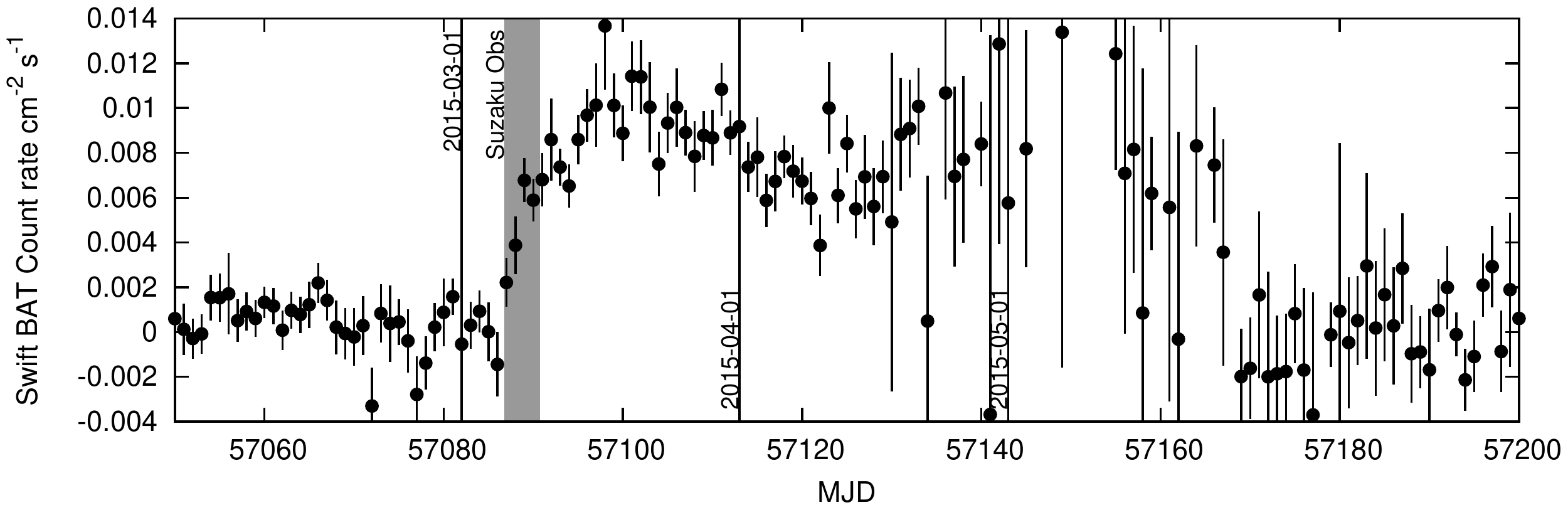}
\caption{
Light curve of {\it Swift}/BAT $15-50$~keV count rate obtained from
the {\it Swift}/BAT Hard X-ray Transient Monitor website.
{\it Suzaku} observation period is shown as shaded region.}
\label{fig:swiftLC}
\end{figure*}

On March 6.84 UTC of 2015 (MJD 57087.84), GK Per was found to have transited into 
the dwarf nova outburst state \citep{wilberetal2015}.
\suzaku X-ray satellite was coincidently observing GK Per as a scheduled
observation from 2015-03-05 18:38:56 UTC (MJD 57086.8)
to 2015-03-09 18:30:11 UTC (MJD 57090.8)
with net exposures of 182.9~ks with the X-ray Imaging Spectrometer 
(XIS; \citealt{koyamaetal2007xis}) and 134.8~ks with the Hard X-ray Detector
(HXD; \citealt{takahashietal2007,kokubunetal2007}).
Table \ref{tab:obslog} summarizes the observation log.
Note that due to the degraded power generation capability of the spacecraft, 
only one XIS unit among four, namely XIS3, was operational during this observation.
In addition, the HXD was operational while the solar paddles were illuminated by the sun, and turned off in the eclipse by the earth.
As shown in Figure \ref{fig:swiftLC}, the \suzaku observation covered
a transition from quiescent to outburst, allowing us to trace temporal variation of X-ray emission; 
the figure utilizes Swift/BAT transient monitor results provided by the Swift/BAT team \citep{krimmetal2013}.

Besides the March 2015 observation, \suzaku has been pointed to the
target twice in February 2009 and August 2014 when the target was in
the quiescent state in optical (Table \ref{tab:obslog}).
Data from these observations are also analyzed in the present report. 
We remark the following two points on these observations;
(1) although three XIS units and the HXD were operational in February 2009,
the source flux was too low to be detected by the HXD,
(2) in August 2014, three XIS units were operational but the HXD
was accidentally left turned off by the cancellation of spacecraft contact operations
due to typhoon interference.

The March 2015 data, together with those from the 2009 and 2014 observations, 
were processed and analyzed using HEASOFT 6.15 maintained and released
by the HEASARC at NASA Goddard Space Flight Center. 
We followed the standard data extraction and event selection procedures of {\it Suzaku}.
The background signals of the XIS was estimated and subtracted using data 
detected in the outer region of the imaging area around the source image.
The HXD/PIN non-X-ray background was subtracted
using the official background model \citep{fukazawaetal2009} released by the \suzaku HXD team 
\footnote{https://heasarc.gsfc.nasa.gov/docs/suzaku/analysis/pinbgd.html}.
The barycentric correction was applied to arrival time of each photon recorded by the XIS and the HXD.

\begin{table*}
\begin{center}
\begin{minipage}{140mm}
\caption{Log of GK Per observations by {\it Suzaku}.}
\label{tab:obslog}
\begin{tabular}{cccccccc}
\hline
Observation ID & Start Date/Time & Exposure$^{a}$ & Aim point & \# of XIS$^{b}$ & Count rate$^{c}$\\
\hline
403081010 & 2009-02-13T10:35:48 & 30.4 & HXD & 3 & $0.25$ \\ 
409018010 & 2014-08-07T17:01:21 & 114.5 & XIS & 3 & $0.89$ \\ 
409018020 & 2015-03-05T18:38:56 & 182.9 & XIS & 1 & $3.08$ \\ 
\hline
\end{tabular}\\
$^{a}$Effective exposure of the XIS in ks.
$^{b}$The number of the XIS unit(s) operational in the observation.
$^{c}$Average $0.5-10$~keV count rates recorded by the XIS FI sensors
in units of counts~s$^{-1}$~XIS$^{-1}$. Statistical errors are less than 0.01 counts~s$^{-1}$.
Note that in the HXD-nominal aim point observation, where a target
point-like source is offset by 3.5~arcmin from the XIS-nominal aim point,
results in $\sim15-20\%$ smaller effective area due to the energy-dependent
vignetting effect of the X-ray mirror.
\end{minipage}
\end{center}
\end{table*}

\section{Analysis and results}\label{sec:analysis}
\subsection{Temporal analysis}\label{sec:temporalAnalysis}

\subsubsection{Light curves}\label{sec:lc}
\begin{figure*}
\includegraphics[width=0.85\hsize]{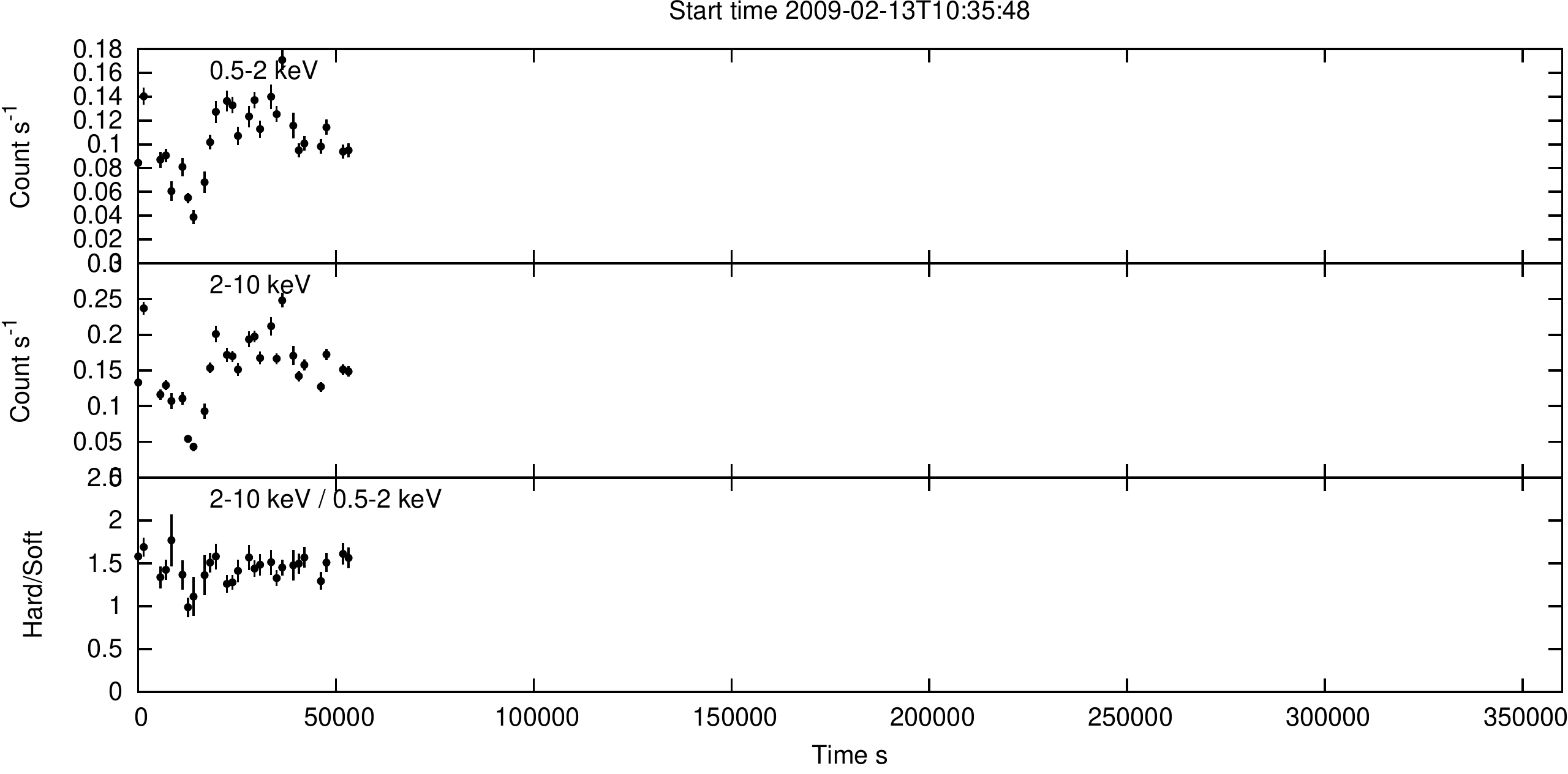}
\includegraphics[width=0.85\hsize]{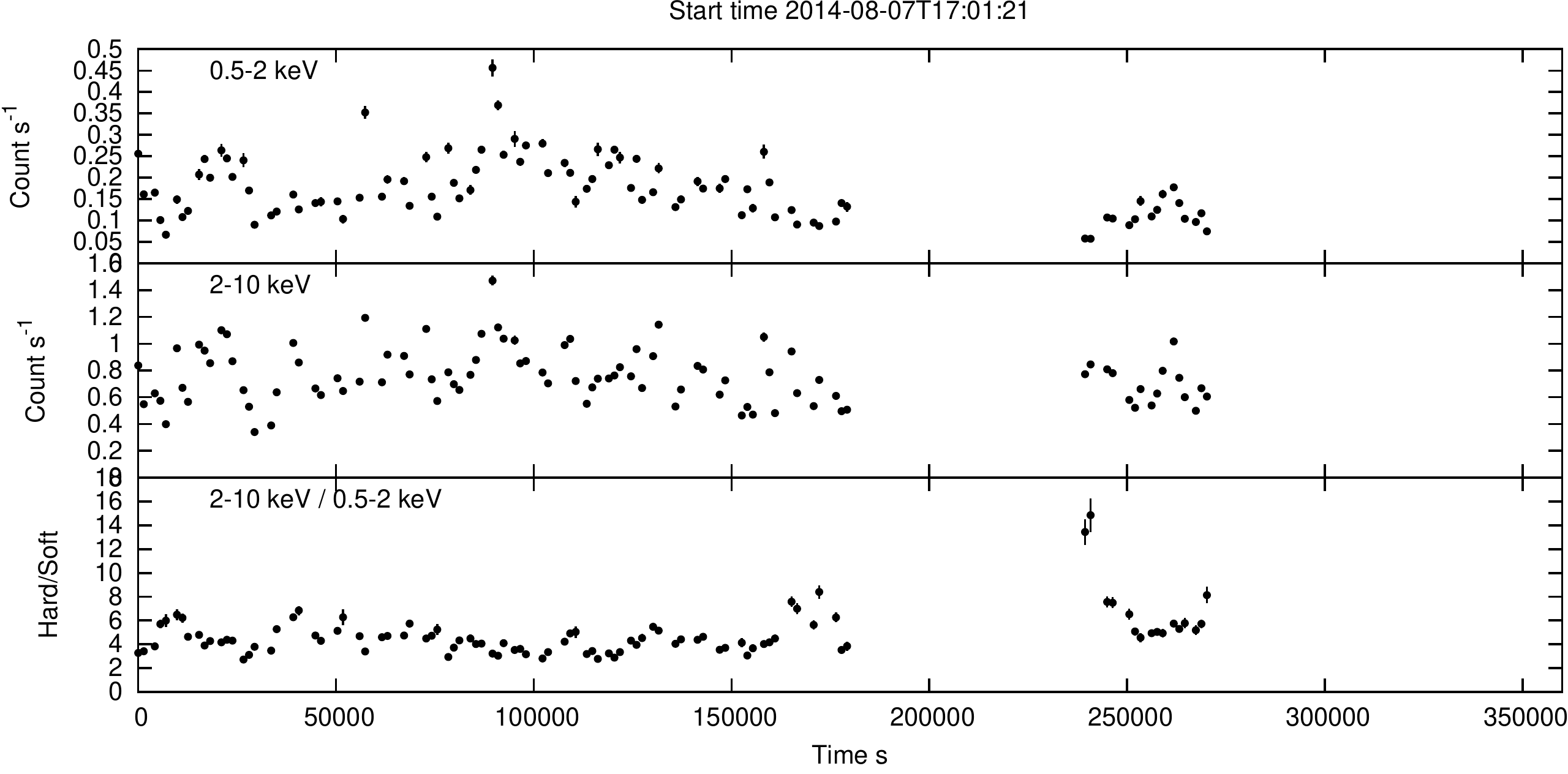}
\includegraphics[width=0.85\hsize]{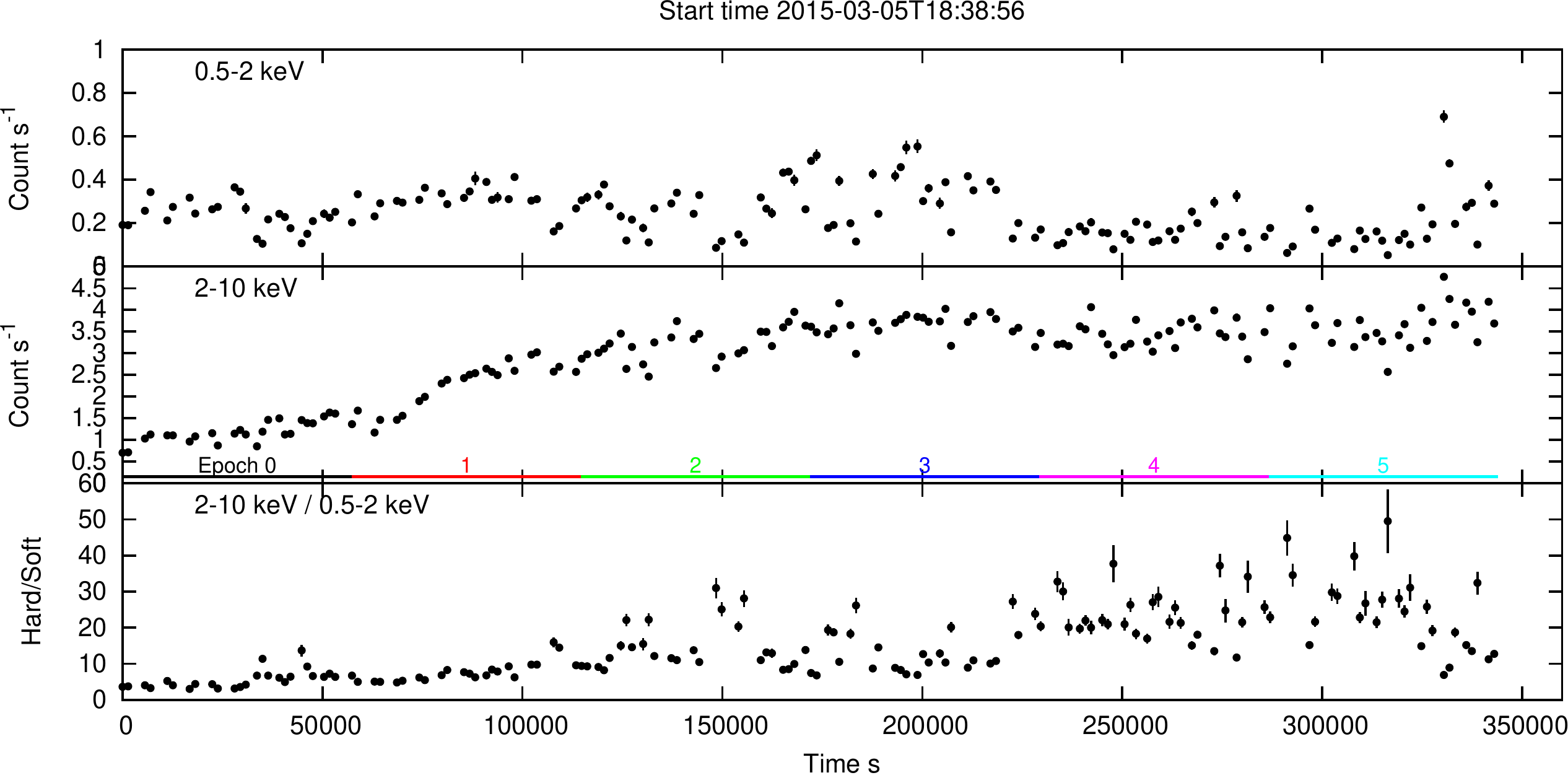}
\caption{
Light curves of XIS count rates of the three observations in February 2009, August 2014, and March 2015. The top, the middle, and the bottom panels in each plot shows a soft-band count rate ($0.5-2$~keV), a hard-band count rate ($2-10$~keV), and the hardness ratio calculated as ($2-10$~keV count rate)/($0.5-2$~keV count rate). Count rate is scaled per XIS unit. Horizontal lines with number labels in the middle panel of the bottom plot represent epochs defined for analyses of the March 2015 data (see text).
}
\label{fig:xisLC}
\end{figure*}

Figure \ref{fig:xisLC} presents light curves of count-rate and hardness ratio.
Apparently, the March 2015 observation nicely covered the quiescent state 
(from 0 to around $5\times10^{4}$~s) and the onset of the dwarf nova outburst.
The count rates of the quiescent period of the March 2015 observation
are almost the same as that of August 2014;
$\sim0.1-0.3$~counts~s$^{-1}$ ($0.5-2$~keV) and 
$\sim1$~counts~s$^{-1}$ ($2-10$~keV).
In the later period of the March 2015 observation, the count rates are rather stable at $\sim0.1-0.2$~counts~s$^{-1}$ ($0.5-2$~keV)
and $\sim3.5-4.5$~counts~s$^{-1}$ ($2-10$~keV).

\begin{figure}
\includegraphics[width=1\hsize]{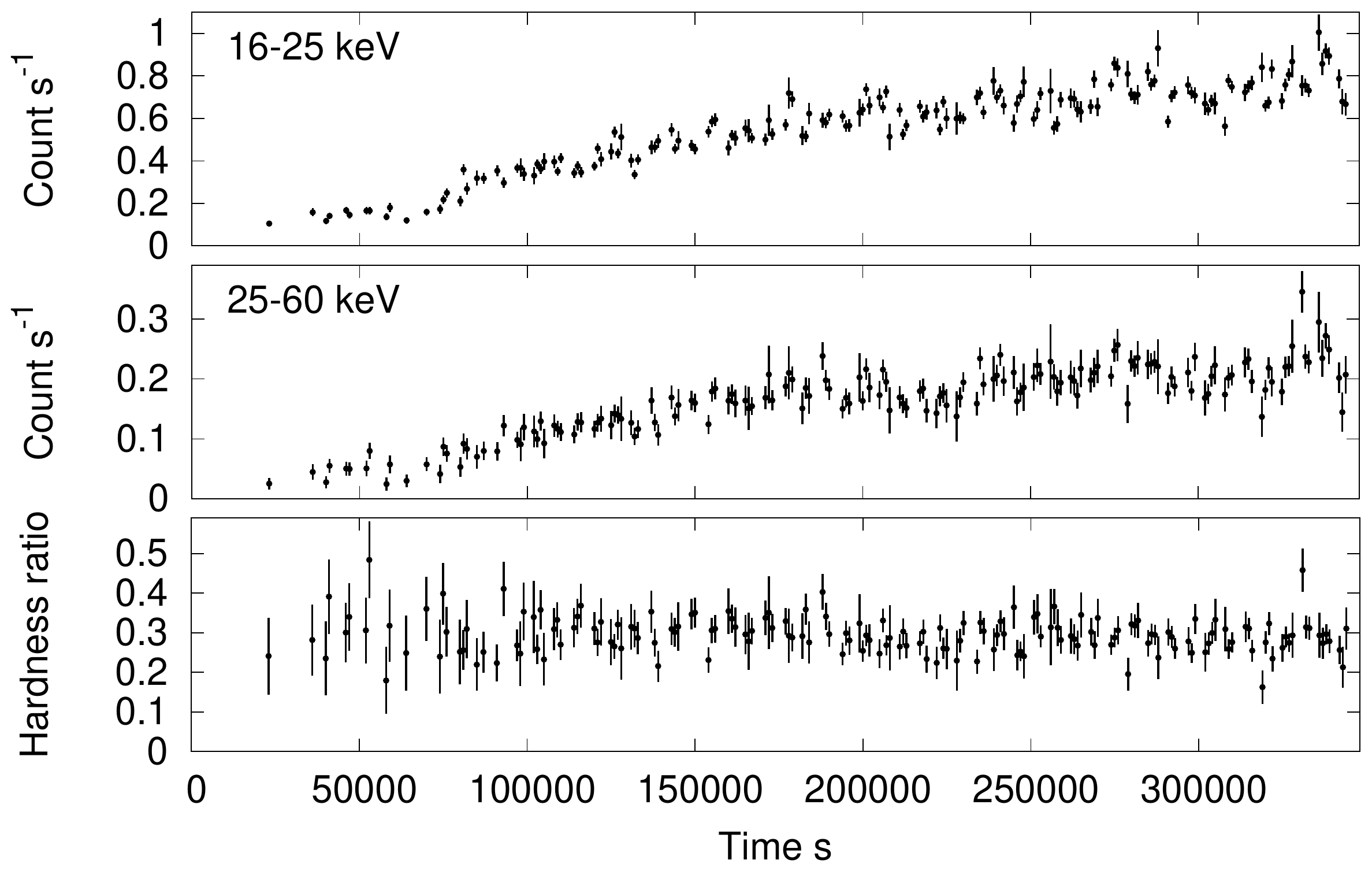}
\caption{
HXD/PIN count-rate light curves in the $16-25$~keV band (top panel) and the $25-60$~keV band (middle panel) of the March 2015 observation. Hardness ratio, calculated as a ratio of count rates ($16-25$~keV)/($25-60$~keV), is plotted in the bottom panel.
}
\label{fig:hxd_lightcurve}
\end{figure}

Hard X-ray ($>16$~keV) light curves of the March 2015 observation are plotted 
in Figure \ref{fig:hxd_lightcurve}.
In contrast to the soft-band signals below 10~keV (Figure \ref{fig:xisLC} bottom three panels),
the hardness ratio calculated in the HXD/PIN energy band is more stable, 
suggesting less change in spectral shape than in the softer energy band.

In the following analyses, to study temporal evolution of spin profiles and 
spectral shapes, we divided the March 2015 data into six equally separated
57320-s time intervals ($\sim0.68$~day) each having an effective XIS exposure of
$\sim30$~ks, and labeled as Epoch 0 to Epoch 5. 
The start time of Epoch 0 was set at the time of the first good time interval of the XIS (2015-03-05 18:55:48.7 UTC; MJD 57086.78954).
During the 2009 and the 2014 observations, GK Per did not
show significant time variability, and therefore, we analyzed their data 
without splitting into shorter time epochs to produce statistically significant data sets.

\begin{figure*}
\includegraphics[height=0.68\hsize]{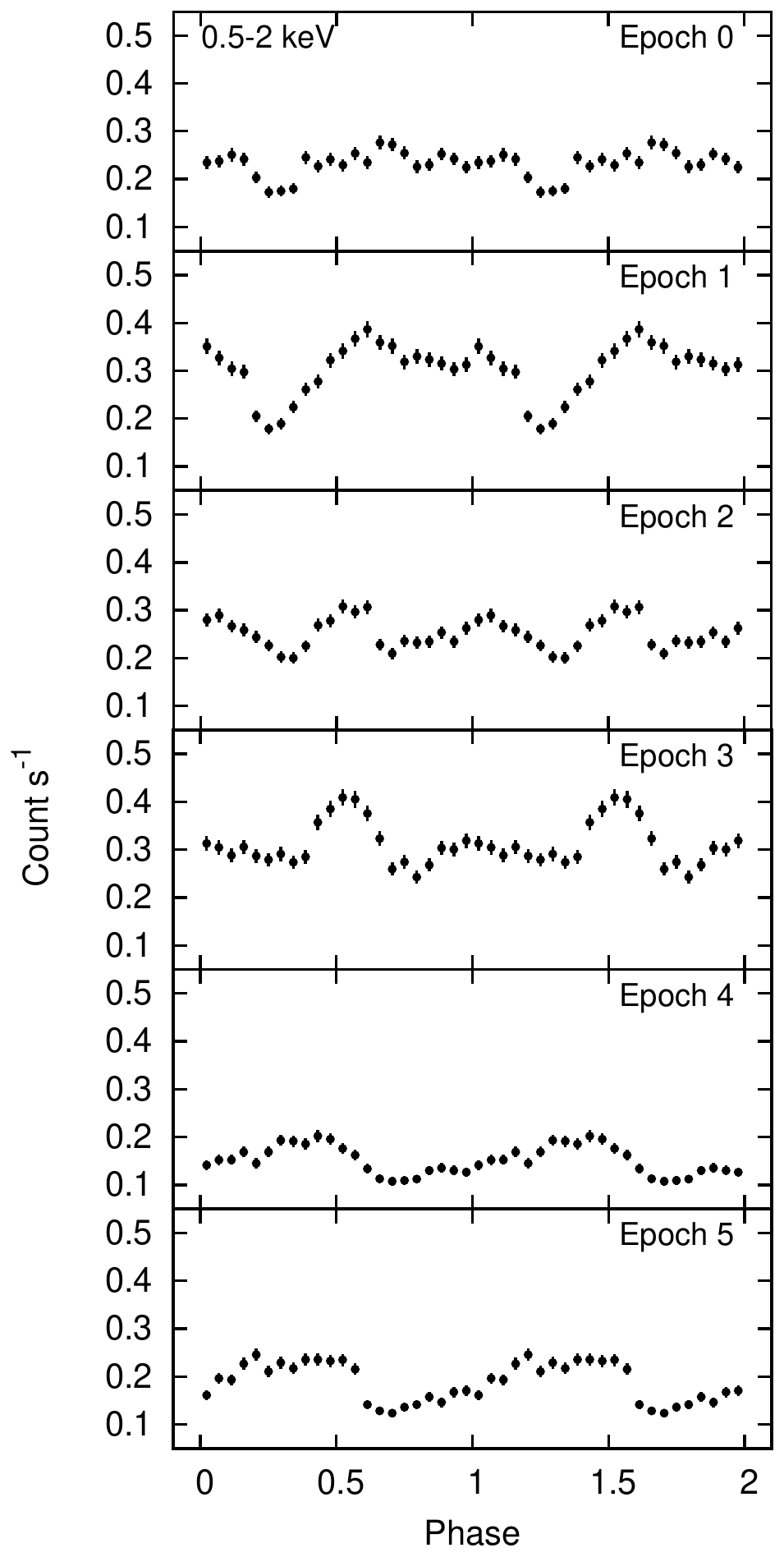}
\includegraphics[height=0.68\hsize]{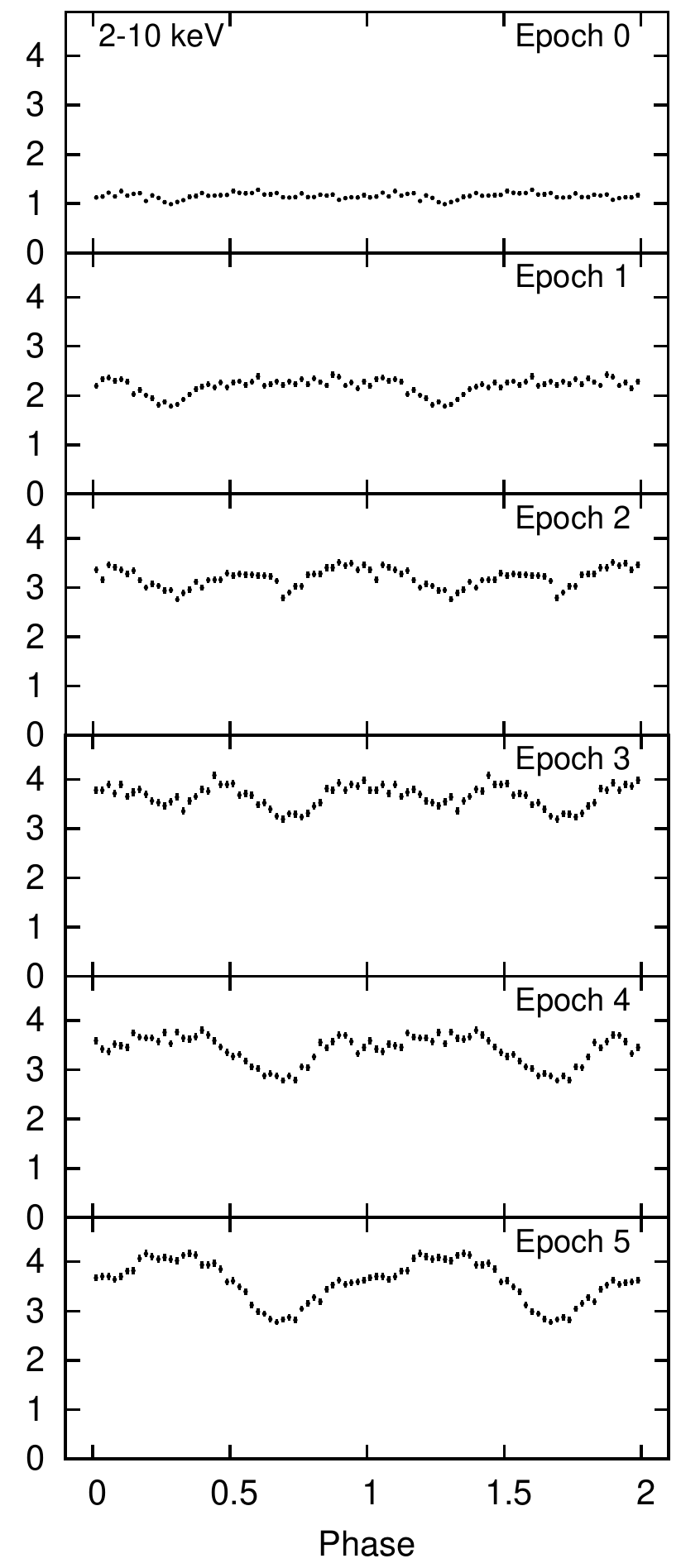}
\includegraphics[height=0.68\hsize]{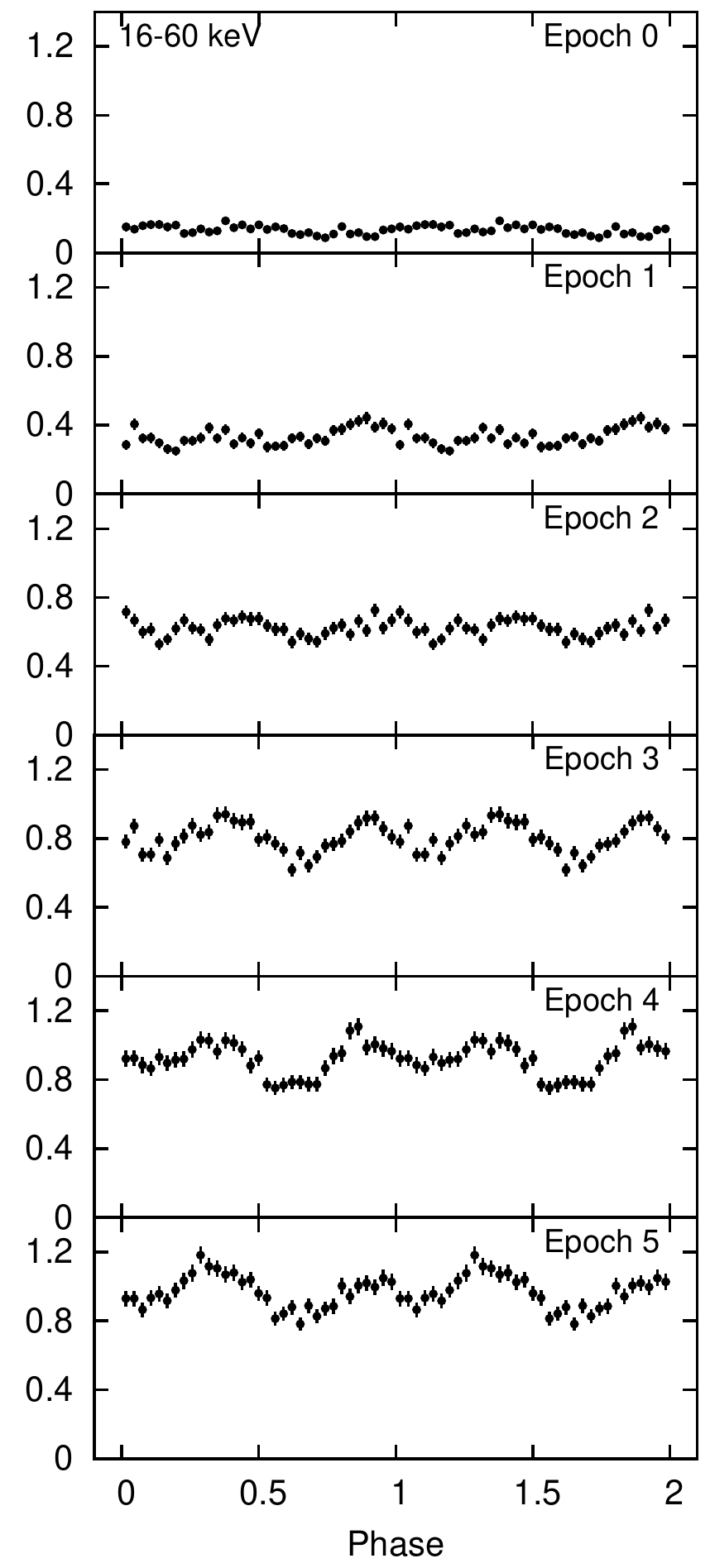}
\caption{
Time evolution of spin-folded light curves of the March 2015 data in three energy bands; 
$0.5-2$, $2-10$, and $16-60$~keV in left, middle, right columns, respectively.
From the top to the bottom, the six panels in each column represent light curves for Epoch 0 to 5.
In each panel, the abscissa covers two spin phases for better visibility.
}
\label{fig:spin_profiles}
\end{figure*}

\subsubsection{Spin profiles}\label{sec:lc}
Using the March 2015 XIS data, we generated 
a power spectrum and a periodgram, and obtained a WD spin period of
$P_{\mathrm{WD}}=351.4\pm0.5$~s.
Based on this, energy-sliced spin-folded light curves were extracted from
the XIS and the HXD/PIN data for each time epoch as shown in Figure \ref{fig:spin_profiles}.
In the folding process, the phase 0 epoch was set at 2015-03-05 00:00:00 UTC (MJD 57086.0), and a spin period of 351.4~s was assumed.
Table \ref{tab:xis_folded_lc} summarizes the mean count rate and the modulation factor
calculated as the full amplitude divided by the mean count rate.

In the figure, a complex time evolution of the spin profiles is seen; for example,
the spin minimum phase shifts from phase $0.2-0.3$ in Epochs $0-2$ to $0.6-0.7$ 
in Epochs $4-5$, and the pulse shape significantly changes epoch by epoch.
As previously reported for the outburst state \citep{watsonetal1985,nortonetal1988,evansetal2009}, 
the overall pulse profile becomes single peaked, especially in the later epochs. 
Throughout the epochs, the modulation factors are higher in the soft band than 
those in the hard band.
The hard X-ray folded light curve also modulates with a modulation factor of $\sim40\%$
(or a min/max ratio of $\sim0.7$) in outburst.

\begin{table*}
\begin{center}
\begin{minipage}{160mm}
\caption{Mean count rate and modulation factor of the spin-folded light curves of the March 2015 outburst. Mean count rates of the 2009 and the 2014 data are also listed for comparison.}
\label{tab:xis_folded_lc}
\begin{tabular}{cccccccccc}
\hline
 & \multicolumn{2}{c}{$0.5-2$ keV} &  & \multicolumn{2}{c}{$2-10$ keV} &  & \multicolumn{2}{c}{$16-60$ keV}\\ \cline{2-3} \cline{5-6} \cline{8-9}
         & Mean$^{a}$ & MF$^{b}$ & & Mean$^{a}$ & MF$^{b}$ & & Mean$^{a}$ & MF$^{b}$\\
\hline
2009 & $0.104\pm0.002$ & $-$ & & $0.151\pm0.002$ & $-$ & & N/A$^{c}$ & $-$ \\
2014 & $0.162\pm0.001$ & $-$ & & $0.730\pm0.002$ & $-$ & & N/A$^{c}$ & $-$ \\
2015\\
Epoch 0 &  $0.232\pm0.003$ & $44.8\pm3.9$  & &$1.155\pm0.006$ & $25.3\pm2.5$  & &$0.133\pm0.004$ & $74.3\pm12.5$ \\
Epoch 1 &  $0.302\pm0.003$ & $68.8\pm3.7$  & &$2.188\pm0.009$ & $29.4\pm1.9$  & &$0.335\pm0.006$ & $57.8\pm6.8$ \\
Epoch 2 &  $0.252\pm0.003$ & $42.7\pm3.8$  & &$3.195\pm0.010$ & $23.7\pm1.5$  & &$0.624\pm0.007$ & $31.7\pm4.3$ \\
Epoch 3 &  $0.311\pm0.003$ & $53.5\pm3.5$  & &$3.653\pm0.012$ & $24.3\pm1.5$  & &$0.805\pm0.007$ & $40.0\pm3.7$ \\
Epoch 4 &  $0.152\pm0.002$ & $62.3\pm4.6$  & &$3.404\pm0.011$ & $30.0\pm1.5$  & &$0.920\pm0.008$ & $38.9\pm3.6$ \\
Epoch 5 &  $0.189\pm0.002$ & $64.8\pm4.2$  & &$3.583\pm0.011$ & $38.9\pm1.4$  & &$0.968\pm0.008$ & $41.2\pm3.5$ \\

\hline
\end{tabular}
\\
$^{a}$In units of counts~s$^{-1} \mathrm{XIS}^{-1}$.
$^{b}$Modulation factor in \%. See text for definition.\\
$^{c}$GK Per flux was below the HXD sensitivity, or the HXD was not operational (see \S\ref{sec:obs}).
\end{minipage}
\end{center}
\end{table*}

\subsection{Spectral analysis}\label{sec:spectral}

\subsubsection{Spectral evolution in the soft energy band}\label{sec:spectral_xis}
Figure \ref{fig:xis_123} compares time-average spectra of the three observations.
Obviously, the source is more luminous in outburst than in quiescence showing about 2 orders of magnitude increase from February 2009 in $>7$~keV. Among the quiescent states, however, detected fluxes differ by a factor of $>5$ (see $2-10$~keV fluxes in Table \ref{tab:xis_nH}), implying a temporal fluctuation of mass accretion rate even during the quiescence. Based on the intensity-scaled spectra in the right of Figure \ref{fig:xis_123}, a major spectral change is observed below 4~keV, and above Fe K line complex ($>7$~keV), the spectra well overlap each other.

For the outburst data, we also extracted time-resolved spectra for the six epochs as presented in Figure \ref{fig:xis_3rd_spectra_all_intervals}. The shape and the normalization of the black spectrum (Epoch 0) almost matches that of the August 2014 time-average spectrum. 
Based on this, we consider that the source was in the quiescent state in Epoch 0, and then a transition to the dwarf nova outburst took place during Epoch 1 where steady increases of the $2-10$~keV and the $16-60$~keV count rates were observed (Figures \ref{fig:xisLC} and \ref{fig:hxd_lightcurve}). The right panel of Figure \ref{fig:xis_3rd_spectra_all_intervals} plots intensity-scaled spectra of individual epochs. As the dwarf nova outburst proceeds, soft X-ray photons below $\sim4$~keV drastically decreased. 
This is a firm indication of an increased absorption column density during the outburst.

In the outburst phase, for example in the most absorbed spectrum of Epoch 5, we detected $0.027\pm0.001$~counts~s$^{-1}$ in the $0.5-1$~keV band, and this rate is 2 or 3 times higher than one expected from heavily ($N_\mathrm{H}>10^{22}$~cm$^{-2}$) absorbed post-shock plasma emission (i.e. extrapolation of $E>2$~keV emission to lower energy band). 
Based on the previous grating observation results \citep{vrielmannetal2005}, we consider that this emission is mostly a sum of line emissions whose origin has not been clearly identified. A detailed spectral modeling of this emission is very challenging with the limited energy resolution of the XIS; we tried to fit this component with multiple gaussian components or phenomenological continuum represented by black body emission but no acceptable fit was obtained. Therefore, in the following spectral fittings, we ignore energies below 2~keV to avoid a complication caused by this soft emission. 

\subsubsection{Absorption column density}\label{sec:xis_nH}
As can be seen from Figures \ref{fig:xis_123} and \ref{fig:xis_3rd_spectra_all_intervals},
absorption of soft X-ray photons is highly variable even in the quiescent state, and obviously during outburst.
The absorption is thought to be mostly intrinsic to GK Per caused by cool pre-shock accreting gas rather than by interstellar medium. To examine an absorption column density increase, we fitted the XIS spectra using a phenomenological model consisting of a bremsstrahlung continuum subject to absorption. The Fe K emission line region, $5.5-7.2$~keV, was ignored to simplify the fitting model. In Epochs $2-5$, a single-absorption model did not reproduce the data, and therefore we introduced a partially-covering absorber. 
Fits are all acceptable, and results are summarized in Table \ref{tab:xis_nH}. The overall absorption column density increases by a factor of $\sim2.4$; from $1.82^{+0.15}_{-0.15}\times10^{22}~\cmsq$ (Epoch 0) to the highest $4.31^{+0.42}_{-0.56}\times10^{22}~\cmsq$ (Epoch 4). The $7.2-10$~keV flux increases by a factor of $\sim4.8$; from 1.35 (Epoch 0) to $6.46\times10^{-11}~\ergcms$ (Epoch 5). Higher partial-covering column densities $\sim13-16\times10^{22}~\cmsq$ was obtained in later epochs and we consider that spin modulation of the line-of-sight column density of this dense absorber creates the hard X-ray modulation detected by the HXD (Figure \ref{fig:spin_profiles}).
Although the face values of bremsstrahlung temperature vary from $19.2^{+7.4}_{-4.5}$~keV (Feb 2009) to $64.4^{+20.8}_{-14.5}$~keV (Epoch 1 of Mar 2015), this is not considered as a real variation because temperatures cannot be accurately determined solely from the XIS data when thermal emission has a temperature higher $kT\sim10$~keV (see \S\ref{sec:spectral_hxd}).

\begin{table*}
\begin{center}
\begin{minipage}{14cm}
\caption{Empirical spectral modeling of the $2-10$~keV continuum emission using a bremsstrahlung continuum and three gaussian components.}
\label{tab:xis_nH}
\begin{tabular}{cccccccc}
\hline
	& $N_{\mathrm{H}}$	& $N^{\mathrm{PC}}_{\mathrm{H}}$$^{a}$	& $kT$ & C.F.$^{b}$ & \multicolumn{2}{c}{Flux$^{c}$} & $\chi^2_{\nu}$~(N.D.F.)$^{d}$ \\
	& $10^{22}~\cmsq$	& $10^{22}~\cmsq$	& keV & & $2-10~\mathrm{keV}$ & $7.2-10~\mathrm{keV}$  \\
\hline
2009 & $0.24^{+0.23}_{-0.22}$ & $-$ & $19.2^{+7.4}_{-4.5}$ & $-$ & 0.72 & 0.18 & 1.26(137)\\
2014 & $1.59^{+0.06}_{-0.06}$ & $-$ & $32.0^{+3.0}_{-2.6}$ & $-$ & 3.1 & 0.93 & 1.25(390)\\
2015\\
Epoch 0 & $1.82^{+0.15}_{-0.15}$ & $-$ & $29.4^{+7.1}_{-5.0}$ & $-$ & 4.4 & 1.35 & $0.97(426)$ \\
Epoch 1 & $2.50^{+0.12}_{-0.12}$ & $-$ & $64.4^{+20.8}_{-14.5}$ & $-$ & 8.9 & 2.98 & $0.99(426)$ \\
Epoch 2 & $2.99^{+0.40}_{-0.68}$ & $13.20^{+6.37}_{-4.72}$ & $39.9^{+19.5}_{-11.1}$ & $0.42^{+0.11}_{-0.06}$ & 13.9 & 5.03 & $1.13(424)$ \\
Epoch 3 & $2.99^{+0.36}_{-0.52}$ & $16.00^{+6.40}_{-5.20}$ & $47.1^{+31.2}_{-15.3}$ & $0.45^{+0.06}_{-0.05}$ & 16.1 & 5.95 & $1.18(424)$ \\
Epoch 4 & $4.31^{+0.42}_{-0.56}$ & $15.05^{+4.37}_{-3.57}$ & $62.5^{+44.3}_{-21.7}$ & $0.53^{+0.06}_{-0.05}$ & 15.7 & 6.24 & $1.00(424)$ \\
Epoch 5 & $3.81^{+0.38}_{-0.47}$ & $16.48^{+3.58}_{-3.04}$ & $43.7^{+22.0}_{-11.8}$ & $0.59^{+0.04}_{-0.04}$ & 16.4 & 6.46 & $1.10(424)$ \\

\hline
\end{tabular}\\
$^{a}$Column density of the partial-covering absorption applied to Epoch 2-5 of the 2015 data.\\
$^{b}$Covering fraction of the partial-covering absorber.\\
$^{c}$Model flux in the $2-10$~keV and the $7.2-10$~keV bands in units of $10^{-11}~\ergcms$. The former does not include $5.5-7.2$~keV flux.\\
$^{d}$Reduced $\chi^2$ and the number of degree of freedom.
\end{minipage}
\end{center}
\end{table*}

\subsubsection{Fe K$\alpha$ lines}\label{sec:xis_fe_line}
\begin{figure}
\includegraphics[width=0.95\hsize]{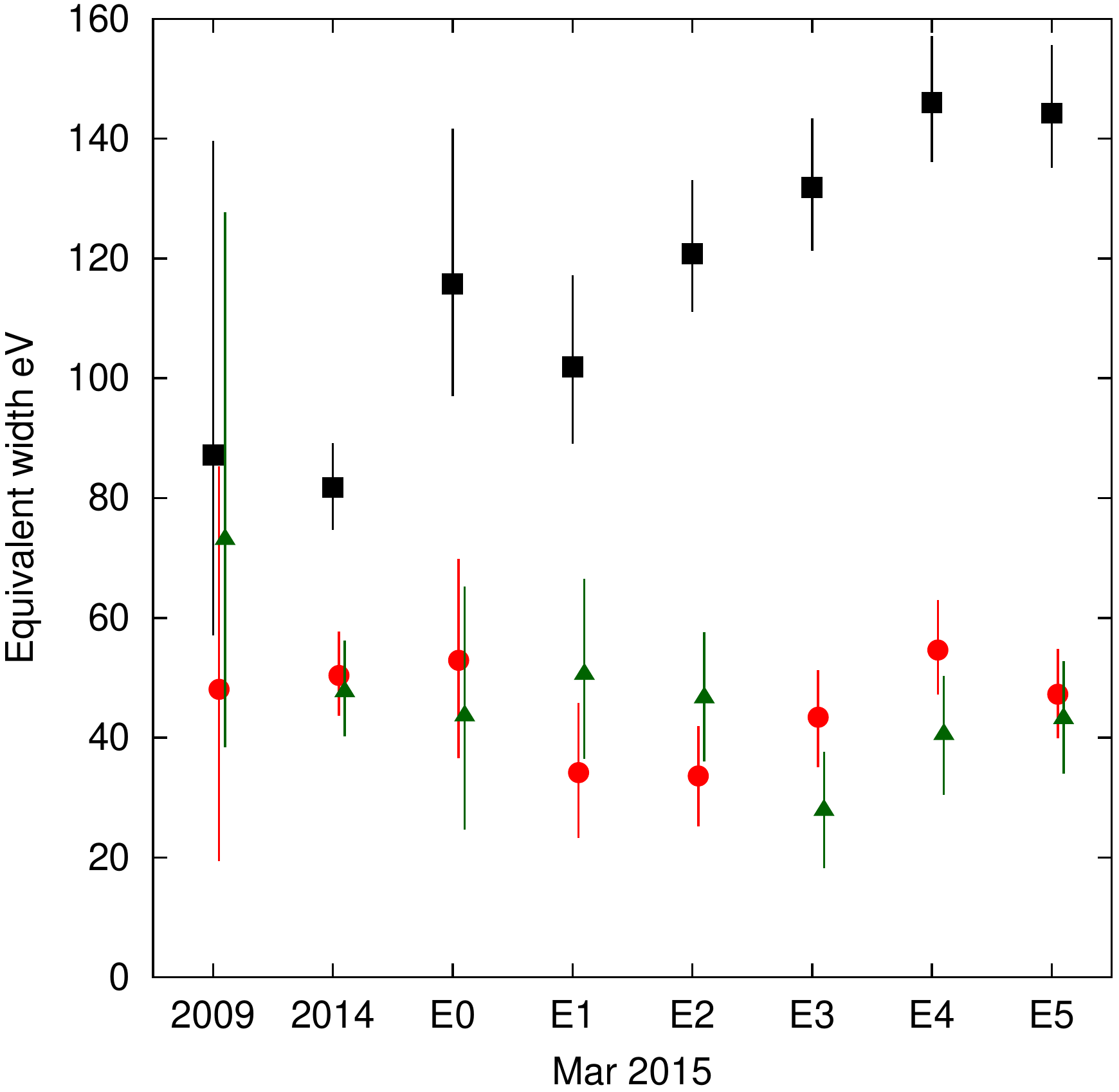}
\caption{
Fe K line equivalent widths derived by an empirical model fitting.
Equivalent widths of the fluorescent, He-like, and H-like Fe K$\alpha$
lines are represented in black, red, and green, respectively.
E0$-$E5 denotes Epoch $0-5$ of the March 2015 observation.
}
\label{fig:fe_line}
\end{figure}

To examine possible time variability of Fe K$\alpha$ line intensities, we also fitted $4-9$ keV narrow-band spectra using an empirical model consisting of a power-law continuum and three gaussians all subject to single absorption. Centroid energies and widths of the three lines were fixed at canonical values (6.40, 6.65, and 6.97~keV) and 0 (i.e. no broadening), respectively, so as to better constrain equivalent widths. The best-fit model functions are plotted in Figure \ref{fig:fe_line_fit} together with the data. Fits were generally acceptable, and we obtained line equivalent widths as listed in Table \ref{tab:fe_line} and plotted in Figure \ref{fig:fe_line}. 

\begin{figure*}
\includegraphics[width=0.48\hsize]{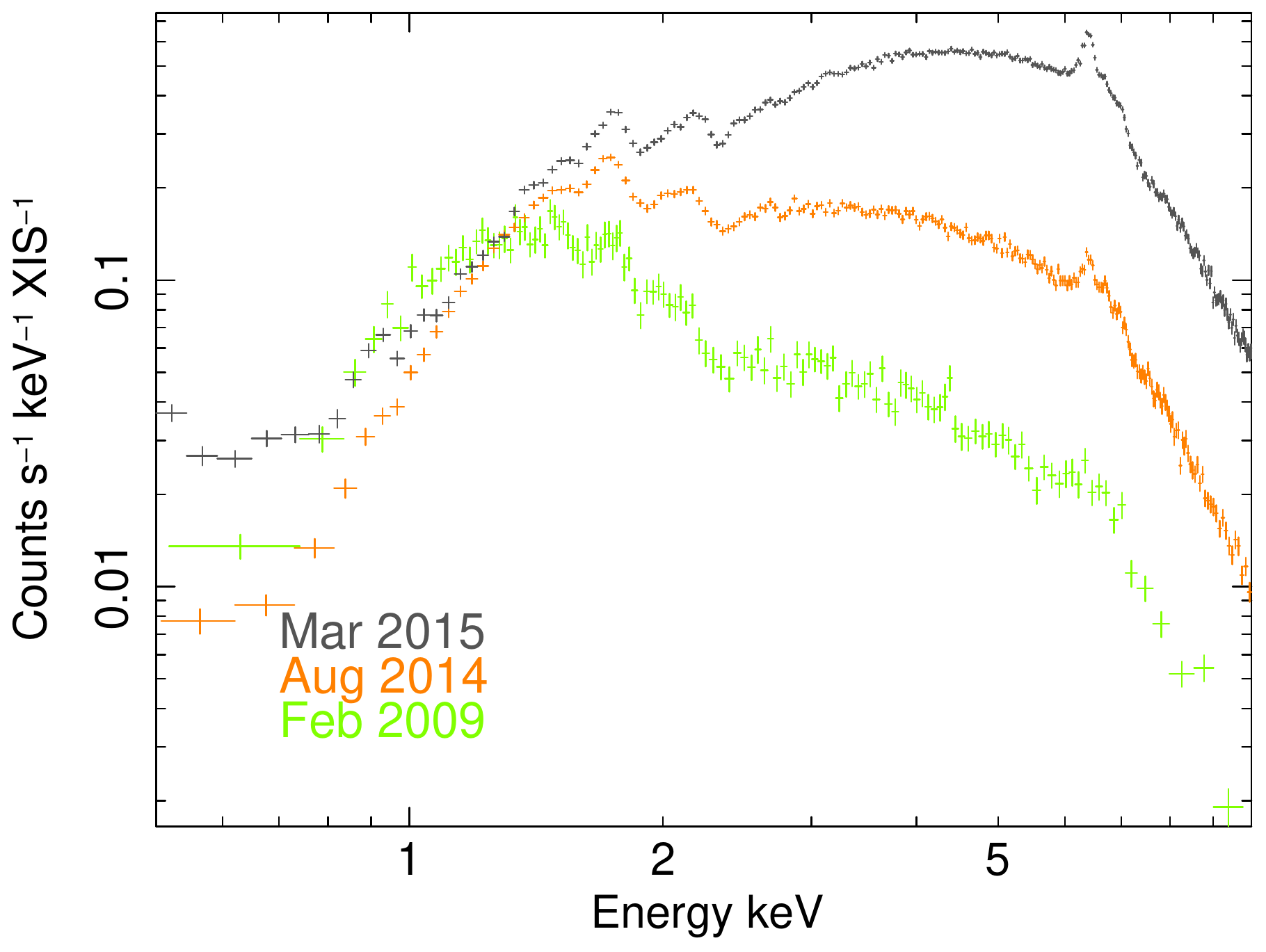}
\includegraphics[width=0.48\hsize]{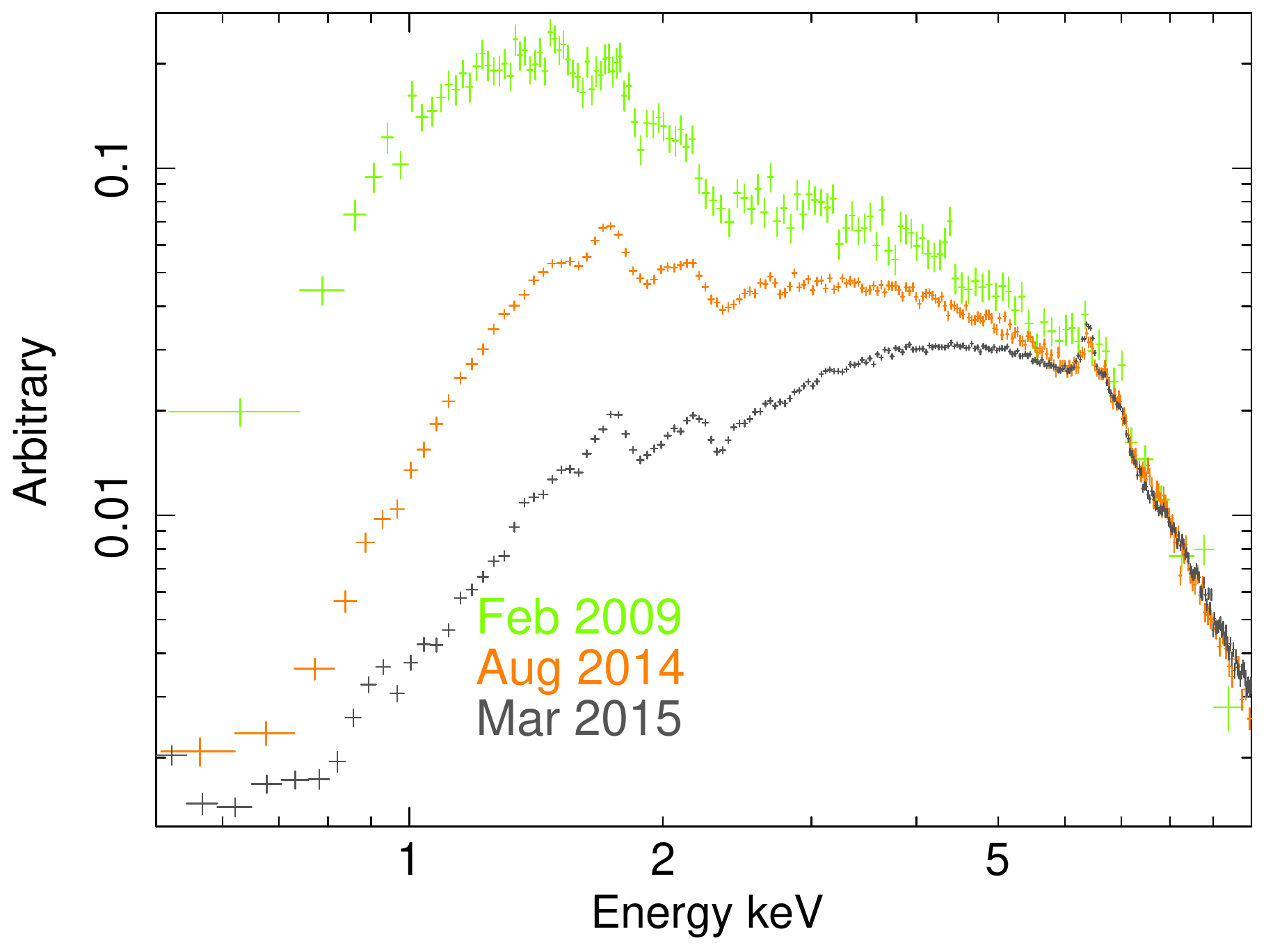}
\caption{
Left: Time-average XIS spectra of the three observations;
green, orange, and dark grey spectra corresponds to observations in February 2009, August 2014, and March 2015, respectively. Ordinate is shown after scaling the number of XIS units operated in each observation.
Right: The same spectra as those in the left panel but vertically scaled for the $7-10$~keV band count rate to match each other; scaling factors are 1, 0.15, and 0.06 for spectra of February 2009, August 2014, and March 2015, respectively.
}
\label{fig:xis_123}
\end{figure*}

\begin{figure*}
\includegraphics[width=0.48\hsize]{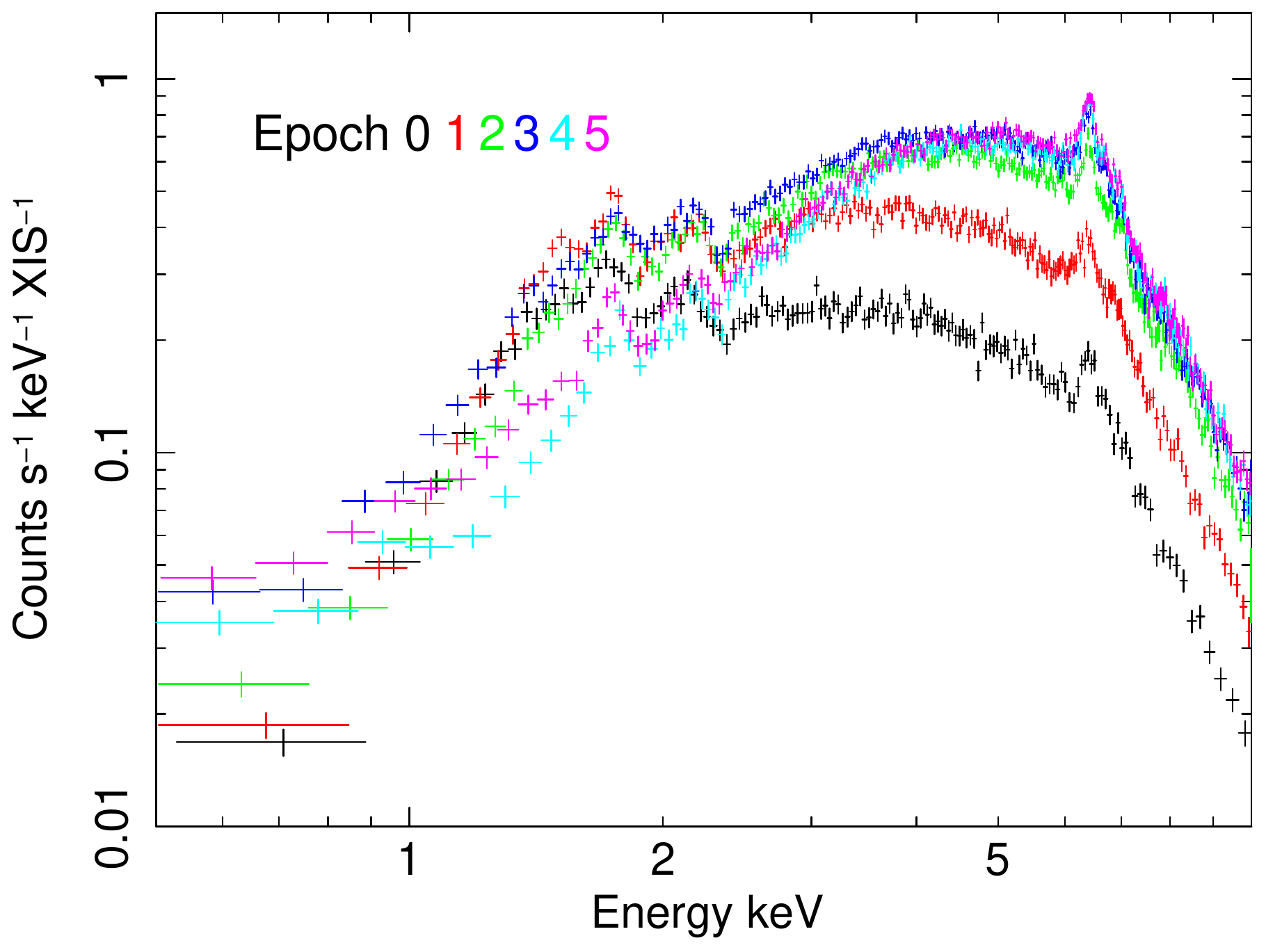}
\includegraphics[width=0.48\hsize]{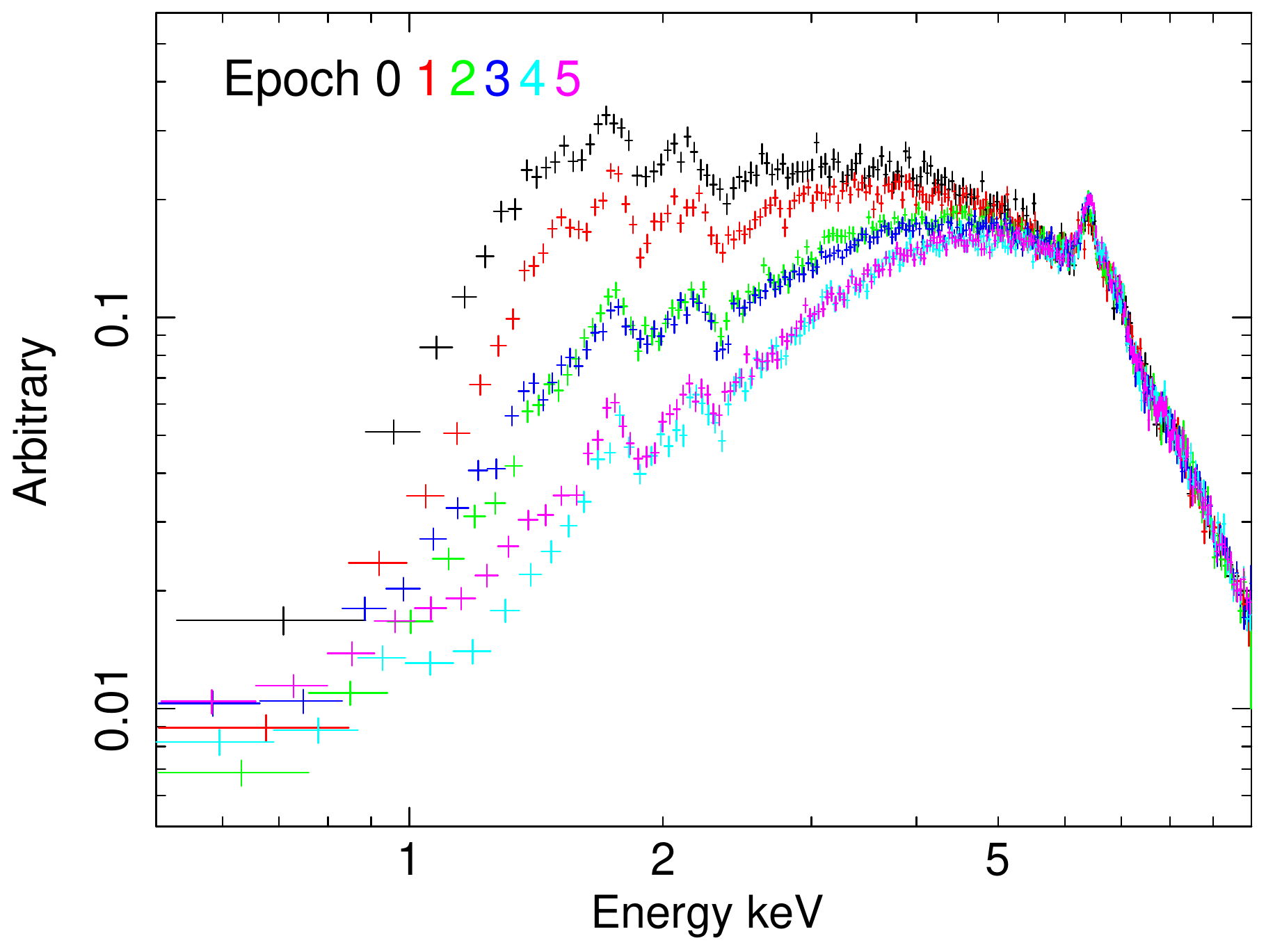}
\caption{
Left: XIS spectra of individual epochs of the March 2015 observation.
Right: The same spectra as those in the left panel but vertically scaled for the $7-10$~keV band count rate to match each other; scaling factors are 1, 0.48, 0.29, 0.24, 0.23, and 0.23 for Epochs $0-5$, respectively.
}
\label{fig:xis_3rd_spectra_all_intervals}
\end{figure*}

\begin{figure}
\includegraphics[width=1\hsize]{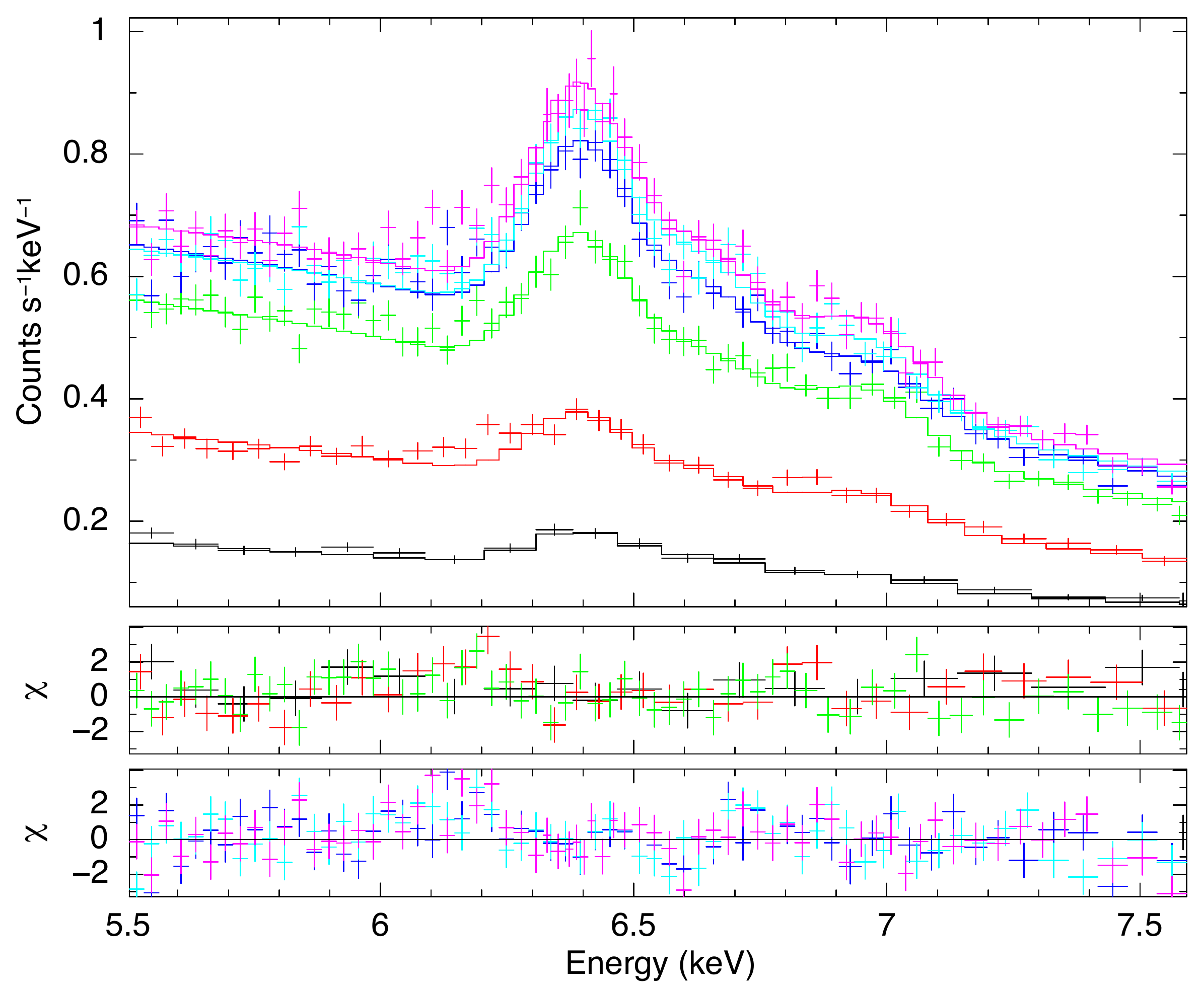}
\caption{
Fe K line fit with three gaussian components and a power-law continuum (top panel).
For better visibility, fit residuals of Epoch 0-2 and 3-5 are separately plotted in the middle and the bottom panels, respectively. Color coding is the same as that of Figure \ref{fig:xis_3rd_spectra_all_intervals}.
}
\label{fig:fe_line_fit}
\end{figure}

The fluorescent Fe K$\alpha$ line increased its equivalent width from $\sim80$~eV in quiescent to $\sim140$~eV in outburst. The equivalent widths of the He-like and H-like Fe K$\alpha$ lines are in the range of $30-50$~eV, and are consistent with being constant throughout the present data set; this is qualitatively consistent with the fact that the continuum spectral shape, in the $7-10$~keV, are very similar in all data sets (Figure \ref{fig:xis_123} and \ref{fig:xis_3rd_spectra_all_intervals}) indicating little change in plasma temperature.


\begin{table}
\begin{center}
\caption{Fe K$\alpha$ line equivalent widths.}
\label{tab:fe_line}
\begin{tabular}{cccc}
\hline
      & Fe I & Fe XXV & Fe XXVI\\
      & eV & eV & eV \\
\hline
2009 & $87^{+52}_{-30}$ & $48^{+37}_{-29}$ & $73^{+55}_{-35}$ \\
2014 & $82^{+7}_{-7}$ & $50^{+7}_{-7}$ & $48^{+9}_{-7}$ \\
2015 \\
Epoch 0 & $116^{+26}_{-19}$ & $53^{+17}_{-16}$ & $44^{+22}_{-19}$ \\
Epoch 1 & $102^{+15}_{-13}$ & $34^{+12}_{-11}$ & $51^{+16}_{-14}$ \\
Epoch 2 & $121^{+12}_{-10}$ & $34^{+8}_{-8}$ & $47^{+11}_{-11}$ \\
Epoch 3 & $132^{+12}_{-11}$ & $43^{+8}_{-8}$ & $28^{+10}_{-10}$ \\
Epoch 4 & $146^{+11}_{-10}$ & $55^{+8}_{-7}$ & $40^{+10}_{-10}$ \\
Epoch 5 & $144^{+11}_{-9}$ & $47^{+8}_{-7}$ & $43^{+10}_{-9}$ \\

\hline
\end{tabular}

\end{center}
\end{table}

\subsubsection{Hard X-ray spectra}\label{sec:spectral_hxd}
\begin{figure}
\includegraphics[width=1\hsize]{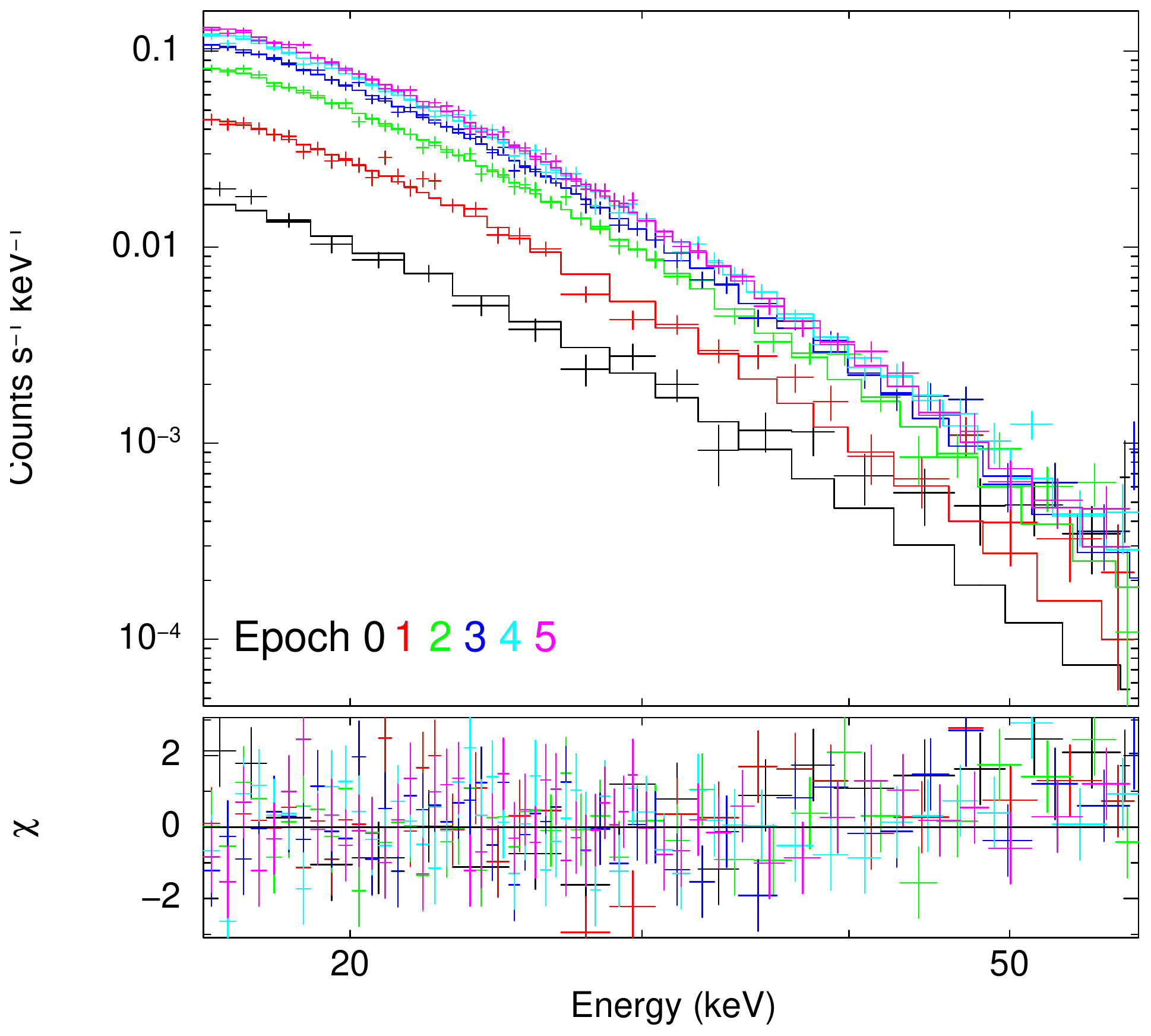}
\caption{
Upper panel: Time-divided HXD/PIN spectra of GK Per taken in the third observation.
Crosses are the background-subtracted signals, and the solid lines represent the best-fit bremsstrahlung model functions.
Lower panel: Fit residual in terms of $\chi$ (data$-$model divided by statistical error).
}
\label{fig:hxd_fit_brems}
\end{figure}

To characterize the hard-band spectrum during the transition to the outburst state,
we extracted epoch-separated HXD/PIN spectra, and fitted with two phenomenological models;
(1) a power-law model and (2) a thermal bremsstrahlung model. 
The fit range was chosen to be $16-60$~keV (the same as the light-curve extraction energy range).
The extracted spectra and the best-fit
model functions are shown in Figure \ref{fig:hxd_fit_brems}, and the best-fit parameters
are listed in Table \ref{table:hxd_fit_pow_brems}.
The best-fit photon indices and temperatures do not show large variation across epochs, 
and are almost consistent each other within statistical fitting errors; $\Gamma=2.2^{+0.3}_{-0.3} - 2.4^{+0.1}_{-0.1}$ and $kT=24.6^{+1.5}_{-1.4} - 31.5^{+15.5}_{-8.5}$~keV.
This is another result supporting non-variability of the intrinsic spectral shape of emission from the post-shock accretion region throughout the quiescence and the early phase of the outburst. In Epoch 0, an apparent deviation of data and the best-fit model curve in the $40-60$~keV band is noticeable in Figure \ref{fig:hxd_fit_brems}. 
This excess cannot be solely explained by the inaccurate background modeling as one might consider, because even in Epoch 0 (quiescent state), the $40-60$-keV band count rate ($0.01$~counts~s$^-1$) is an order of magnitude higher than that of the background model accuracy in the same energy band. Although we do not fully understand the origin of the excess, exclusion of the energy band from the fitting results similar best-fit parameters as listed in Table \ref{table:hxd_fit_pow_brems}, and therefore we leave the excess as is in the present analysis. 

\begin{table*}
\begin{center}
\begin{minipage}{140mm}
\caption{HXD/PIN spectral fit result of the March 2015 data using a power-law model or a bremsstrahlung model.}
\label{table:hxd_fit_pow_brems}
\begin{tabular}{cccccccc}
\hline
      & \multicolumn{7}{c}{Parameter} \\ \cline{2-8}
      & \multicolumn{3}{c}{Power Law} & & \multicolumn{3}{c}{Bremsstrahlung} \\ \cline{2-4} \cline{6-8}
Epoch & $\Gamma$$^{a}$ & $\chi^{2}_{\nu}$(N.D.F.)$^{b}$ & $F_{16-60}$$^{c}$ & & $kT$$^{d}$ & $\chi^{2}_{\nu}$(N.D.F.)$^{b}$ & $F_{16-60}$$^{c}$ \\
\hline
0 & $2.2^{+0.3}_{-0.3}$ & 0.95(71) & 0.74 & & $31.5^{+15.5}_{-8.5}$ & 1.02(71) & 0.70\\
1 & $2.3^{+0.1}_{-0.1}$ & 1.17(78) & 1.79 & & $25.7^{+3.9}_{-3.1}$ & 1.18(78) & 1.67\\
2 & $2.3^{+0.1}_{-0.1}$ & 0.89(81) & 3.35 & & $27.2^{+2.5}_{-2.2}$ & 0.79(81) & 3.16\\
3 & $2.3^{+0.1}_{-0.1}$ & 1.15(83) & 4.32 & & $26.0^{+1.9}_{-1.7}$ & 0.93(83) & 4.06\\
4 & $2.3^{+0.1}_{-0.1}$ & 1.18(85) & 5.01 & & $27.2^{+1.9}_{-1.7}$ & 0.87(85) & 4.73\\
5 & $2.4^{+0.1}_{-0.1}$ & 1.46(84) & 5.11 & & $24.6^{+1.5}_{-1.4}$ & 1.13(84) & 4.81\\
\hline
\end{tabular}

\\
$^{a}$Power-law photon index.
$^{b}$Reduced $\chi^2$ and the number of degree of freedom.\\
$^{c}$Model predicted $16-60$~keV flux in units of $10^{-11}~\ergcms$. 
$^{d}$Temperature in keV.
\end{minipage}
\end{center}
\end{table*}

\subsubsection{Broad-band spectral interpretation}\label{sec:spectral_broad}
We also fitted broad-band combined spectra of the XIS and the HXD/PIN
in the $2-60$~keV. The broad-band model consists of an isobaric cooling-flow
plasma emission model (developed for galaxy cluster studies; \citealt{mushotzkyzymkowiak1988})
subject to a single-column photo absorption and a partial-covering 
photo absorption to mimic multi-column absorption \citep{donemagdziarz1998} and spin modulation of
column density. Reflection on freshly accreted material on the WD surface
is modeled using the convolution model \verb|reflect| by \citealt{magdziarzandzdziarski1995}.
A gaussian is utilized to model the fluorescent Fe K$\alpha$ line emitted
by the reflection. The composite model can be denoted in Xspec as
\verb|phabs|$\times$\verb|pcfabs|$\times($\verb|reflect|$\times$\verb|mkcflow|$+$
\verb|gaus|$)$.
The cooling-flow model is a good approximation to the X-ray emission from 
the intermediate polars' post-shock accretion region (\citealt{mukaietal2015,hayashietal2011}), and in the present analysis, 
we utilized the model aiming at assessing a potential change of the maximum temperature in the post-shock accretion region. 
The reflector's covering area against the irradiator (i.e. hot plasma in the post-shock accretion region) and relative normalization of the reflection are fixed at $2\pi$ and $1$, respectively, assuming that the shock height is much smaller than the radius of the WD (for example, see \citealt{yuasaetal2010}). Since the angle dependent reflection was hardly detected in the present data, we tentatively fixed the reflection angle, or an angle between the reflecting surface normal and the line of sight, at the default value of $\cos\mu=0.45$. Fe abundance of the cooling-flow model and the reflection model are tied and allowed to freely vary.

\begin{figure*}
\includegraphics[width=0.65\hsize]{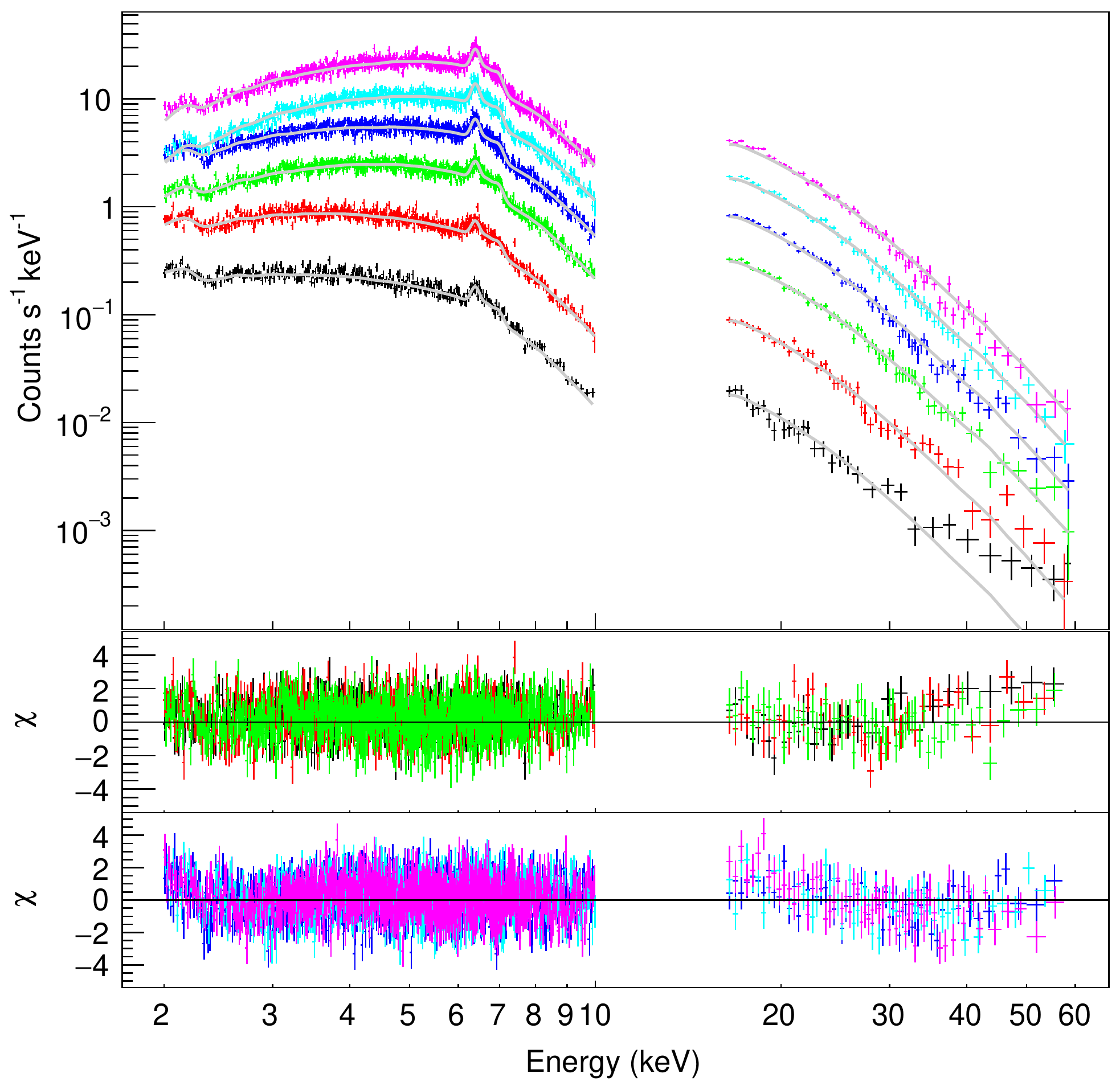}
\caption{
Broad-band fit result of Epochs 0-5 of the March 2015 data. Color coding is the same as that of Figure \ref{fig:xis_3rd_spectra_all_intervals}.
The top panel shows the data and the best-fit models with each data set shifted upward by a step of a factor of 2 for better visibility. The middle and the bottom panels show fit residuals of Epochs $0-2$ and Epochs $3-5$, respectively.
}
\label{fig:broad_band_fit}
\end{figure*}

The spectra of individual epochs were separately fitted, and the best-fit results were obtained as plotted in Figure \ref{fig:broad_band_fit}. Table \ref{tab:broad_band_fit} summarizes the best-fit parameters. The fits are generally good with the null hypothesis probabilities larger than $1\%$ all except for Epoch 5 ($0.1\%$). The overall absorption column density, the first column of Table \ref{tab:broad_band_fit}, are almost consistent with those derived from the XIS-only fitting (Table \ref{tab:xis_nH}) within the statistical fitting errors, and the tendency of increase over time is common between the two modelings. The apparent increase of the covering fraction may suggest more frequent obscuration of the post-shock plasma by a pre-shock cold gas in later phases of the outburst. The face values of the highest temperature of the cooling-flow model increase through Epoch 0 to Epoch 2, and are consistent with being constant (within errors) in Epochs 2-5.

The maximum temperature reported by the cooling-flow model is generally higher than one obtained with the PIN-only bremsstrahlung fitting (Table 5) in each epoch. This is an expected result because the bremsstrahlung model only provides a representative temperature that simulates a spectrum from a post-shock region that has an emissivity gradient, being weighted by lower-temperature component which has higher emissivity (due to increased density) and photon statistics. When the PIN spectra are fitted with a cooling-flow model, the maximum temperatures similar to those from the XIS+PIN fitting (Table \ref{tab:broad_band_fit}) are obtained; e.g. $kT_{\mathrm{max}}=47.2^{+19.6}_{-11.6}$~keV and $49.0^{+3.5}_{-2.8}$~keV for Epoch 0 and 5, respectively.

\begin{table*}
\begin{center}
\begin{minipage}{14cm}
\caption{Result of broad-band spectral fit using an partially-absorbed multi-temperature emission model convolved with a reflection model.}
\label{tab:broad_band_fit}
\begin{tabular}{cccccccc}
\hline
Epoch & $N_{\mathrm{H}}$$^{\mathrm{a}}$   & $N^{\mathrm{PC}}_{\mathrm{H}}$$^{\mathrm{b}}$ & C.F.$^{\mathrm{c}}$ & $kT_{\mathrm{max}}$$^{\mathrm{d}}$ & $Z_{\mathrm{Fe}}$$^{\mathrm{e}}$ & $\chi^2_{\nu}$~(N.D.F.)$^{f}$\\
      & $10^{22}$~cm$^{-2}$ & $10^{22}$~cm$^{-2}$ &      & keV          & $Z_\odot$          & \\
\hline
0 & $1.79^{+0.16}_{-0.09}$ & $62.2^{+31.5}_{-47.9}$ & $0.17^{+0.17}_{-0.15}$ & $41.8^{+9.8}_{-9.8}$ & $0.10^{+0.06}_{-0.05}$ & $0.94$ (1308)\\
1 & $2.60^{+0.13}_{-0.15}$ & $64.3^{+15.9}_{-20.1}$ & $0.30^{+0.09}_{-0.10}$ & $45.1^{+8.6}_{-6.8}$ & $0.09^{+0.04}_{-0.03}$ & $1.04$ (1782)\\
2 & $3.53^{+0.22}_{-0.29}$ & $23.6^{+8.3}_{-6.7}$ & $0.35^{+0.03}_{-0.03}$ & $59.3^{+4.7}_{-5.0}$ & $0.12^{+0.03}_{-0.02}$ & $0.99$ (1980)\\
3 & $3.75^{+0.16}_{-0.17}$ & $40.7^{+7.4}_{-6.8}$ & $0.45^{+0.03}_{-0.03}$ & $55.2^{+5.1}_{-5.0}$ & $0.08^{+0.02}_{-0.02}$ & $1.06$ (2007)\\
4 & $5.50^{+0.22}_{-0.23}$ & $41.2^{+7.6}_{-6.8}$ & $0.49^{+0.03}_{-0.02}$ & $59.4^{+5.9}_{-5.7}$ & $0.10^{+0.02}_{-0.02}$ & $1.03$ (2011)\\
5 & $4.88^{+0.23}_{-0.23}$ & $34.6^{+5.9}_{-4.9}$ & $0.53^{+0.02}_{-0.02}$ & $59.7^{+4.6}_{-4.8}$ & $0.11^{+0.02}_{-0.02}$ & $1.10$ (2056)\\
\hline
\end{tabular}\\
$^{a}$Column density of the single-column absorption.\\
$^{b}$Column density of the partial-covering absorption.\\
$^{c}$Covering fraction.\\
$^{d}$The highest temperature of the cooling-flow model.\\
$^{e}$Fe abundance.\\
$^{f}$Reduced $\chi^2$ and the number of degree of freedom.\\
\end{minipage}
\end{center}
\end{table*}

\section{Discussion}\label{sec:discussion}

\subsection{Accretion geometry in quiescent and outburst states}\label{sec:disc_acc_geometry}


The broad-band spectral modeling in outburst showed that the maximum plasma temperature in the post-shock accretion region is almost constant while the hard X-ray flux increased by a factor of $\sim6-7$ from Epoch 0 to Epoch 5 (Tables \ref{tab:xis_nH} and \ref{tab:broad_band_fit}). In the same time period, the absorption column density has increased about three times (Table \ref{tab:broad_band_fit}).
We consider that these increases reflect an enhanced mass accretion rate, and in the present discussion, we adopt the flux enhancement factor as representative of mass transfer rate, because the line-of-sight column density may not have linear dependence on an actual mass accretion rate due to geometrical effect (e.g. \citealt{vrielmannetal2005}). Therefore, it is assumed that about $6-7$ times more mass transfer took place in Epoch 5 than that in Epoch 0. 

In magnetic CVs, it is thought that accreting matter is transformed to magnetically-channeled accretion from the accretion disk at a radius where the magnetic pressure balances the gas pressure.
This radius is the so-called magnetospheric radius $R_{\mathrm{A}}$ \citep{warner1995,franketal2002accretionpower}, and scales with the mass accretion rate $\dot{m}$ as $R_{\mathrm{A}}\propto\dot{m}^{-2/7}$.
Since in the current observation period, $\dot{m}$ increased by a factor of $6-7$, and using this scaling, the magnetospheric radius in quiescence ($R_{\mathrm{A,qui}}$) and outburst ($R_{\mathrm{A,out}}$) can be related as $R_{\mathrm{A,out}}\sim0.6~R_{\mathrm{A,qui}}$.

If we assume a WD mass of $M_\mathrm{WD}=0.87~M_\odot$ \citep{morales-ruedaetal2002gkper} and that the quiescent magnetospheric radius should be matched to the co-rotation radius $R_\mathrm{A,qui}=R_\Omega=11.1~R_\mathrm{WD}$ (for $P_{\mathrm{WD}}=351.4$~s), we obtain $R_{\mathrm{A,out}}\sim0.6~R_{\mathrm{A,qui}}=6.6~R_\mathrm{WD}$ when disk-magnetosphere interaction is in equilibrium at an accretion rate of the outburst.
Following the calculations in \citet{suleimanovetal2005}, the maximum plasma temperature in the post-shock accretion region should decrease as
\begin{eqnarray*}
T^{\mathrm{out}}_\mathrm{max} \simeq \left( \frac{1-R_\mathrm{WD}/R_\mathrm{A,out}}{1-R_\mathrm{WD}/R_\mathrm{A,qui}} \right) T^{\mathrm{qui}}_\mathrm{max}
\sim0.93~T^{\mathrm{qui}}_\mathrm{max}.
\end{eqnarray*}
This difference is relatively small compared to the $\sim10\%$ statistical fitting errors associated to the maximum temperature obtained from the broad-band fitting (Table \ref{tab:broad_band_fit}), and therefore it is difficult to definitively discuss the possibility of the temporal variation of the maximum temperature due to the shrunk inner disk radius. However, based on the result that the face values of the maximum temperature do not monotonically decrease throughout the epochs, we may consider that, during the \suzaku observation covering about three days after the onset of the outburst, the inner disk radius was rather stable at the similar radius as $R_{\mathrm{A,qui}}$, and probably slowly transiting toward a new equilibrium radius at the increased mass accretion rate. In the previous outburst observation of XY Ari, \citet{hellieretal1997} argued that the disk inner edge transition from $\sim9~R_\mathrm{WD}$ to $\sim4~R_\mathrm{WD}$ took place in about 1~day based on consideration that the inner disk that approached to the WD blocked the view of the lower accretion pole resulting large ($\sim\times10$) pulse amplitude in outburst. \citet{hellieretal2004gkper} mentioned a possibility of similar ``blocking of accretion pole by disk'' for the GK Per outburst. However, at least during the early outburst ($1-3$ days from the onset), our result suggests that the radius of the inner edge of the disk is rather constant and the sinusoidal modulation seen in outburst is solely caused by obscuration by accreting pre-shock gas.

\subsection{Fe fluorescence line shape}
A red-ward wing-like structure that has been moderately detected and reported by \citet{hellierandmukai2004fek} using \chandra HETG data in outburst is an interesting topic regarding the Fe fluorescence line. 
The authors considered that the red-ward wing could be a Doppler shifted Fe fluorescence line emitted from the pre-shock absorbing gas moves at velocities close to the free-fall velocity (a few times $1000~\kms$; see also \citealt{hayashietal2011} for a similar study in V1223 Sgr).

In the present study, however, the Fe fluorescence line measured by the XIS can be explained by a single gaussian without any broadening or centroid shift (\S\ref{sec:xis_fe_line}). It will be difficult to detect the reported structure, if exists, due to a degraded CCD energy resolution (FWHM$\sim$200~eV in 2015). In the near future, the {\it ASTRO-H} X-ray microcalorimeter will allow us to measure the line shape in great detail with an energy resolution of FHWM$\sim5$~eV, and to separate the Doppler shifted component from the fluorescence line from the WD surface. If the interpretation of reflection off the pre-shock gas is validated, the Doppler measurement will provide another strong constraint on the accretion geometry in outburst.

\section{Conclusions}
\begin{itemize}
\item \suzaku serendipitously observed an onset of the dwarf nova outburst of the intermediate polar GK Per in March 2015. In the latter part of the observation, the flux reached at about 50\% of the maximum of the present outburst (\S\ref{sec:obs}).
\item The WD spin period of $351.4\pm0.5$~s was clearly detected in outburst, and the hard X-ray signals above 16~keV were also modulated at the same period (\S\ref{sec:lc}) suggesting absorption by a dense obscurer with $N_{\mathrm{H}}>10^{23}~\cmsq$.
\item The Fe K fluorescent line intensity increased from $\sim80$~eV to $\sim140$~eV following the onset of the outburst (\S\ref{sec:xis_fe_line}).
\item Multi-temperature spectral model fit to the time-sliced broad-band spectra in $2-60$~keV revealed no significant change of the maximum temperature of the post-shock accreting flow in outburst (\S\ref{sec:spectral_broad}). This may suggest that, although the mass accretion rate increased by $\sim6-7$ times compared to quiescence, the inner accretion disk radius changed little during the observation until $\sim3$~days after the outburst onset. 
\end{itemize}

\section*{Acknowledgments}
The authors appreciate the \suzaku operation team that made this research possible. 
We also acknowledge the use of public data from the Swift data archive.
T.Y. is supported by the Special Postdoctoral Researchers Program in RIKEN.
This work was partly supported by JSPS KAKENHI Grant Number 15K17668.

\bibliographystyle{mnras}
\bibliography{bibtex_library}

\label{lastpage}

\end{document}